\documentclass[%
 reprint,
 amsmath,amssymb,
 aps,
]{revtex4-2}

\usepackage{graphicx}
\usepackage{dcolumn}
\usepackage{bm}
\usepackage{amsmath}
\usepackage{float}
\usepackage{subfigure}
\usepackage[colorlinks=true,urlcolor=blue,citecolor=black,linkcolor=black]{hyperref}
\usepackage{url}
\begin{document}


\title{Countermeasure for negative impact of practical source in continuous-variable measurement-device-independent quantum key distribution}


\author{Luyu Huang}
\affiliation{State Key Laboratory of Information Photonics and Optical Communications, Beijing University of Posts and Telecommunications, Beijing 100876, People's Republic of China}

\author{Xiangyu Wang}
\email[]{xywang@bupt.edu.cn}
\affiliation{State Key Laboratory of Information Photonics and Optical Communications, Beijing University of Posts and Telecommunications, Beijing 100876, People's Republic of China}

\author{Ziyang Chen}
\email[]{chenziyang@pku.edu.cn}
\affiliation{State Key Laboratory of Advanced Optical Communication Systems and Networks, School of Electronics and Center for Quantum Information Technology, Peking University, Beijing 100871, People's Republic of China}

\author{Yanhao Sun}
\affiliation{State Key Laboratory of Information Photonics and Optical Communications, Beijing University of Posts and Telecommunications, Beijing 100876, People's Republic of China}

\author{Song Yu}
\affiliation{State Key Laboratory of Information Photonics and Optical Communications, Beijing University of Posts and Telecommunications, Beijing 100876, People's Republic of China}

\author{Hong Guo}
\affiliation{State Key Laboratory of Advanced Optical Communication Systems and Networks, School of Electronics and Center for Quantum Information Technology, Peking University, Beijing 100871, People's Republic of China}


\date{\today}

\begin{abstract}
Continuous-variable measurement-device-independent quantum key distribution (CV-MDI QKD) can defend all attacks on the measurement devices fundamentally. Consequently, higher requirements are put forward for the source of CV-MDI QKD system. However, the imperfections of actual source brings practical security risks to the CV-MDI QKD system. Therefore, the characteristics of the realistic source must be controlled in real time to guarantee the practical security of the CV-MDI QKD system. Here we propose a countermeasure for negative impact introduced by the actual source in the CV-MDI QKD system based on one-time-calibration method, not only eliminating the loophole induced from the relative intensity noise (RIN) which is part of the source noise, but also modeling the source noise thus improving the performance. In particular, three cases in terms of whether the preparation noise of the practical sources are defined or not, where only one of the users or both two users operate monitoring on their respective source outputs, are investigated. The simulation results show that the estimated secret key rate without our proposed scheme is about 10.7 times higher than the realistic rate at 18 km transmission distance when the variance of RIN is only 0.4. What's worse, the difference becomes greater and greater with the increase of the variance of RIN. Thus, our proposed scheme makes sense in further completing the practical security of CV-MDI QKD system. In other words, our work enables CV-MDI QKD system not only to resist all attacks against detectors, but also to close the vulnerability caused by the actual source, thus making the scheme closer to practical security.
\end{abstract}


\maketitle

\section{Introduction}

Quantum key distribution (QKD) is a recently-developed technique that allows remote legitimate users, usually called Alice and Bob, to extract secure keys with theoretically unconditional security, which is guaranteed by the principles of quantum mechanics~\cite{Xu_2020_RevModPhys_92_025002,Pirandola_AdvOptPhoton_4_1012}. Generally, QKD protocols can be achieved by two strategies, that is, discrete-variable strategy and continuous-variable (CV) strategy~\cite{Ralph_1999_PhysRevA_61_010303, Grosshans_2003_Nature_421, Weedbrook_2004_PhysRevLett_93_170504}, with respect to the dimension of the Hilbert space where the information is encoded on. Although the latter was proposed later, it has attracted much attention since it adopts standard optical components and more information is carried per symbol ~\cite{Weedbrook_2012_RevModPhys_84_621,Diamanti_2015_Entropy17_096072}. Therefore, it is able to theoretically attain higher secret key rate and have better compatibility with the existing optical network. In addition, CV-QKD is much easier to be integrated on chips~\cite{Zhang_2019_NatPhoton_13_839}. Due to its powerful advantages, significant progress of CV-QKD has been developed in the field of theoretical analysis~\cite{Navascues_2006_PhysRevLett_97_190502, GarciaPatron_2006_PhysRevLett_97_190503, Leverrier_2013_PhysRevLett_110_030502, Leverrier_2015_PhysRevLett_114_070501,Leverrier_2017_PhysRevLett_118_200501, Ghorai_2019_PhysRevX_9_021059, Lin_2019_PhysRevX_9_041064, Chen_2019_Entropy_21_652, Chen_2022_Preprint} as well as experimental implementation~\cite{Grosshans_2003_Nature_421, Qi_2007_PhysRevA_76_052323, Jouguet_2012_NatPhoton_7, Jouguet_2012_OptExp_20_14030, Huang_2016_OptLett_41_3511, Zhang_2019_QuanSciTech_4_035006, Aguado_2019_IEEECommunMag_57_20, Guo_2021_FundamentalResearch_1_96, Huang_2021_NewJPhys_23_113028} up to now.

Generally, CV-QKD imposes some assumptions on the devices in the security proof, which are hard to satisfy in realistic implementation. These gaps between theoretical assumptions and experimental implementation   may lead to security loopholes, which are potentially exploited by Eve to implement various attacks~\cite{Ma_2013_PhysRevA_88_022339, Jouguet_2013_PhysRevA_87_062313, Qin_2016_PhysRevA_94_012325, Huang_2013_PhysRevA_87_062329}. To solve this, some assumptions that the devices are trustworthy are removed to meet the realistic conditions, and various device-independent or partial-device-independent protocols have been proposed, including CV measurement-device-independent (MDI) protocol~\cite{Pirandola_2015_NatPhoton_9_397, Ma_2014_PhysRevA_89_042335, Li_2014_PhysRevA_89_052301}, which can defend all the attacks against measurement device. Due to this great ability, CV-MDI QKD protocol has attained a lot of progress in theory~\cite{Lupo_2018_PhysRevA_97_052327, Chen_2018_PhysRevA_98_012314, Huang_2019_Entropy_21_1100, Bai_2020_QuantumInfProcess_19_53} and has been experimentally demonstrated in lab~\cite{Pirandola_2015_NatPhoton_9_397}, making a common scenario in communication possible where two users can only be connected through an untrusted third party.

The detectors in the measurement side are not easy to attack in the CV-MDI QKD system, and, relatively speaking, the sources are fragile ~\cite{Stiller_2015_CLEO, Zheng_2019_PhysRevA_100_012313, Zheng_2019_OptExp_27_27369, Huang_2016_OptQuantumElectronics_48_430, Ma_2019_PhysLettA_383_126005, Wang_2020_PhysRevA_102_022609} due to the gap between the practical laser and the ideal assumption. Specifically, imperfect modulation and output intensity fluctuation contribute additional noise in the preparation step. Ignoring the preparation noise may open a loophole and thus result in the damage of CV-MDI QKD system. The estimation of the preparation noise is a valid way to determine the characters of the actual states to be propagated. The normal method is to separate a part of the quantum signal before it enters the channel, and the part is then measured using homodyne detection~\cite{Yang_2012_PhysRevA_86_042314}. However, the laser intensity fluctuation contributes the intensity noise, which is frequently characterized by the relative intensity noise (RIN). As an additional component of the total preparation noise other than shot noise and electronic noise, the RIN is usually neglected when considering the source noise or estimating the shot noise, which will result in the inaccuracy of the estimated shot noise unit (SNU), a key normalization parameter in quantifying the measurement results of monitoring. Further, the estimated results in terms of the incorrect monitoring results will deviate from the realistic ones~\cite{Chu_2021_QuantSciTech_6_025912}, resulting in the overestimation of the secret key rate, which leaves a potential security loophole. Besides, the imperfections, including the electronic noise and limited detection efficiency, of the practical detector in the monitoring module is not involved in the previous work~\cite{Yang_2012_PhysRevA_86_042314}.

To solve this, we propose a countermeasure for negative impact of the actual source in CV-MDI QKD system. Considering the imperfections of the detectors in the practical situation and modeling them based on one-time calibration (OTC) method~\cite{Zhang_2020_PhysRevApplied_13_024058}, the proposed scheme enables the users to determine the preparation noise. That is, utilizing our proposed scheme enables the preparation noise to be modeled. Later, the noise is trusted and controlled by the users and the uncertainty between the users is reduced, making the improvement of performance. Besides, with the proposed scheme, the RIN can be involved in the SNU calibration, thus can eliminate the vulnerability of ignoring its existence. Thus, our proposed scheme enables CV-MDI QKD system not only to resist all attacks against detectors, but also to close the vulnerability caused by the actual source, thus making the scheme closer to practical security. In particular, we investigate three cases according to the fragility of the sources (easy or not to be attacked) of the two users. The security analyses of these three cases in finite-size regime are then given in details. In the numerical simulation, the impact of RIN on the performance of CV-MDI QKD in the three cases is firstly investigated. Then, the performance with and without our proposed scheme is compared. Finally, the effect of beamsplitter (BS) transmittance of the monitoring module is simulated for further optimize the performance.

The rest of the paper is arranged as follows: In Sec.~\ref{CV-MDI QKD}, we briefly review the CV-MDI QKD protocol with Gaussian-modulated coherent states, and propose our schemes corresponding to three cases mentioned above in realistic scenario, respectively. In Sec.~\ref{Security analysis}, we show how to derive secret key rates in finite-size regime in three cases. The simulation results and discussion are given in Sec.~\ref{Results}. Finally, we draw the conclusion in Sec.~\ref{Conclusion}.

\section{Countermeasure for realistic source in CV-MDI QKD with RIN involved}\label{CV-MDI QKD}

In this section, we will first review the original CV-MDI QKD protocol with Gaussian-modulated coherent states and the original method to monitor what Alice and Bob really prepare. Then, a scheme for monitoring the states from the actual laser in CV-MDI QKD coupled with OTC method is proposed, where RIN is involved. In particular, three cases in terms of the vulnerability of the source are investigated, where only Alice can estimate her preparation noise, only Bob can estimate his preparation noise, and both users can estimate their preparation noise, respectively.

\subsection{New model to determine the practical source noise of CV-MDI QKD with RIN involved}

In the prepare-and-measure (P\&M) version of the original CV-MDI QKD protocol with coherent states, two remote parties, Alice and Bob, generate a series of coherent states $\left|\alpha\right\rangle$ and $\left|\beta\right\rangle$. Their amplitudes $\alpha$ and $\beta$ satisfy independent identical Gaussian distribution with zero mean and variance of $V = V_{\text{Mod}} + 1$. Then the states $\left|\alpha\right\rangle$ and $\left|\beta\right\rangle$ are sent through two different quantum channels with length $L_{AC}$ and $L_{BC}$ to the untrusted relay, Charlie. After receiving these states, Charlie performs CV Bell detection. Finally, the measurement results are publicly announced, according to which Alice and Bob are allowed to establish correlation on their data. In this structure, Alice and Bob are able to distribute secret key even if Charlie is totally controlled by Eve. In other words, CV-MDI QKD protocols can defend all the attacks against detectors.

However, imperfections of practical source also results in the fragility of CV-MDI QKD system. Thus, determining what Alice and Bob really prepare is significant in the CV-MDI QKD system. The general approach to monitor the source is to separate a part of the light and detect it using homodyne detection (or conjugate homodyne detections) to define the source noise, and the other part of the light is used for key distribution. In particular, the measurement results are determined by the calibrated SNU. In the realistic CV-MDI QKD system, however, the presence of RIN introduced from the laser adds additional component to total noise, and thus affects the evaluated value of SNU. Neglecting RIN will make Alice and Bob underestimate excess noise and overestimate the secret key rate (See Appendix~\ref{AppendixA} in details). In this scenario, Eve can attain secure information and destroy the practical security of CV-MDI QKD system without being noticed.

\begin{figure*}
\includegraphics[width=6 in]{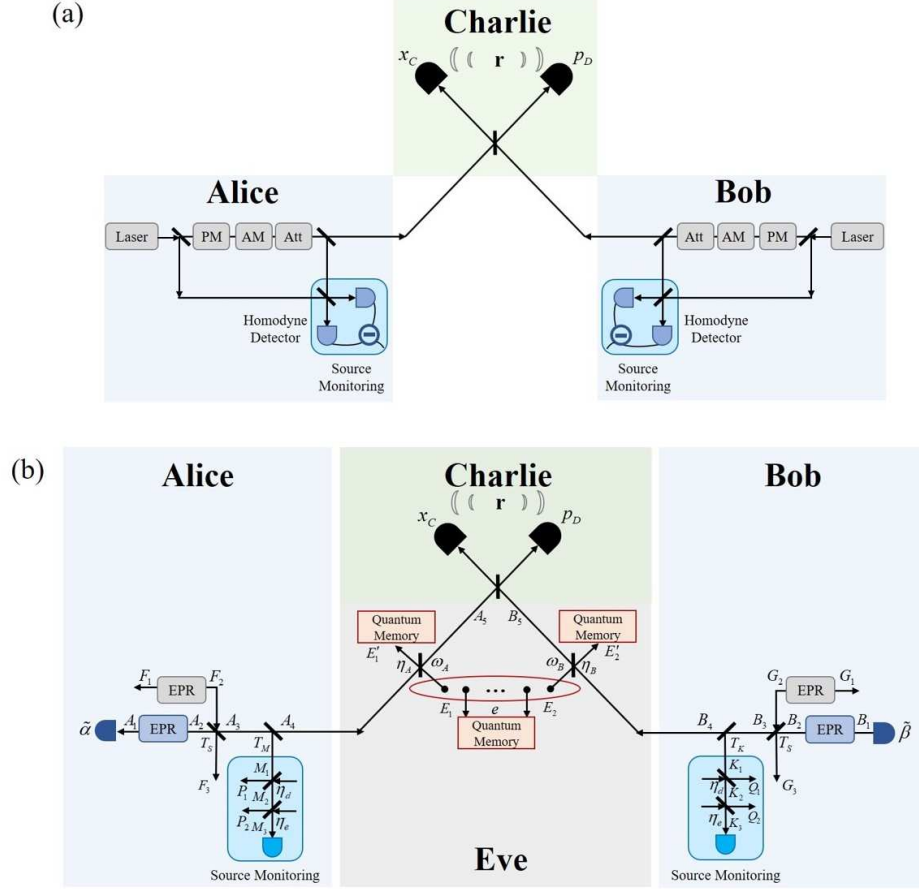}
\caption{The P\&M and EB version of the proposed scheme for CV-MDI QKD. (a) P\&M version. Alice (Bob) modulates the lasers employing the phase modulator (PM) and amplitude modulator (AM) to generate the Gaussian coherent states. Then she (he) separates a part of light to define what the practical prepared states are by homodyne detection. The other part of the light is sent to the untrusted party, Charlie, who performs CV Bell detection on the received states. (b) EB version. The detection efficiency and electronic noise of the homodyne detector in the monitoring module are modeled by two BSs with transmittance $\eta_{\text{d}}$ and $\eta_{\text{e}}$ respectively. Att: optical attenuator.}\label{PM&EBscheme}
\end{figure*}

To avoid the security issue induced from the RIN, here we adopt a countermeasure to eliminate the negative impact induced from RIN with OTC method for CV-MDI QKD system. The P\&M version of our proposed scheme is shown in Fig.~\ref{PM&EBscheme}(a) and conducted as follows: a portion of the light emitted from the laser and the aforementioned quantum signal for monitoring input together into the homodyne detectors, where the former is used for the LO path and the latter is used for the signal path in the monitoring module respectively. Here the variance of the detector's output with the LO switched on is the so-called total noise $V_{\text{tot}}$. Note that in our work, the total noise $V_{\text{tot}}$ is treated as the new SNU $u^{\prime}$. Due to the existence of RIN, we have $u^{\prime}= V_{\text{tot}}=u+V_{\text{el}}+V_{\text{RIN}}$, which has been considered to be $V_{\text{tot}}=u+V_{\text{el}}$ in the previous works~\cite{Jouguet_2012_PhysRevA_86_032309, Lodewyck_2007_PhysRevA_76_042305, Khan_2013_PhysRevA_88_010302}. Note that $u$ is the original SNU, $V_\text{{el}}$ is the variance of electronic noise of the practical detector used for monitoring, and $V_\text{{RIN}}$ is the variance of RIN. In the equivalent entanglement-based (EB) scheme, two independent BSs with transmittance $\eta_{\text{d}}$ and $\eta_{\text{e}}$, are used for modeling the efficiency and electronic noise of the detector, respectively. In particular, the EB version of the CV-MDI QKD protocol with our proposed scheme is revealed in Fig.~\ref{PM&EBscheme}(b), and the establishment of the equivalence between the two versions of the OTC method is reviewed in Appendix~ \ref{AppendixB}.

\subsection{Three cases of the proposed scheme for CV-MDI QKD}

Here three slightly changed schemes corresponding to three cases for CV-MDI QKD system employing the above method are present respectively. Particularly, the three cases are sorted by the fragility of the sources (easy or not to be attacked). That is, only Alice defines her source noise, only Bob defines his source noise, and both users define their source noise, separately. To do this, they conduct the trusted modeling of their source noise so as to control it, which has been considered to be under the control of Eve before. Thus, the three cases mentioned above correspond to the conditions where only the preparation noise in Alice's side is modeled, only the preparation noise in Bob's side is modeled, and both the preparation noise in Alice's and Bob's sides are modeled. For the convenience of security analysis, we give a universal scheme depicted in Fig.~\ref{PM&EBscheme}(b). Here we assume the transmittance of the BS used to separate a beam of light for monitoring is $\eta_{\text M}$, then we can adapt $T_{M(K)}$ to be $\eta_{\text M}$ or 1 to coincide with the three cases.

\subsubsection{Case 1: only Alice defines her prepared states}

In this case, only Alice can estimate her preparation noise due to the monitoring module, while Bob sends the states to Charlie directly. Namely, $T_M=\eta_{\text M}, T_K = 1$. Here defining the preparation noise corresponds to conducting the trusted model of the preparation noise. Since the monitoring module is not used in Bob's side, the preparation noise is impossible to be modeled, correspondingly modes $G_1$, $G_2$ and $G_3$ are not involved. The EB scheme corresponding to this case is described as follows:

\emph{Step 1}. Alice (Bob) prepares an Einstein-Podolsky-Rosen (EPR) state $\rho_{A_1A_2}~(\rho_{B_1B_2})$ with variance of $V = V_{\text{Mod}} + 1$. One mode $A_1$ ($B_1$) of this EPR state is detected by heterodyne detection, resulting in the other mode $A_2$ ($B_2$) being projected onto a coherent state $\left|\alpha\right\rangle$ ($\left|\beta\right\rangle$).

\emph{Step 2}. Assume that noise introduced from the imperfect modulation is modeled by an EPR state with variance of $V_{\text S}$ and a BS with transmittance $T_{\text S}$ at each side~\cite{Usenko_2010_PhysRevA_81_022318} where possible, thus the state that Alice outputs is actually $\rho_{A_1A_3}$ in this case. Since the monitoring module is not set in Bob's side, the preparation noise is impossible to be modeled and the state that Bob outputs keeps $\rho_{B_1B_2}$. Note that if there exists RIN, the total noise variance is $V_{\text S}+V_{\text{RIN}}$ instead of $V_{\text S}$.

\emph{Step 3}. Then mode $A_3$ is injected into a BS with transmittance $T_M$, outputting modes $A_4$ and $M_1$, while mode $B_2$ is injected into a BS with transmittance $T_K=1$, outputting mode $B_4$ (note that $B_4$ is effectively the same as $B_2$ in this case). In the monitoring module in Alice, subsequently, mode $M_1$ passes through the first BS with transmittance $\eta_{\text d}$, outputting modes $M_2$ and $P_1$. Next, $M_2$ passes through the second BS with transmittance $\eta_{\text e}$, outputting modes $M_3$ and $P_2$. Finally, $M_3$ is measured with homodyne detection.

\emph{Step 4}. Modes $A_4$ and $B_4$ are sent through the untrusted channel with transmittance $\eta_A$ and $\eta_B$ to Charlie, where $\eta_A = 10^{-\alpha L_{AC}/10}$ and $\eta_B = 10^{-\alpha L_{BC}/10}$, separately, with the fiber attenuation $\alpha= 0.2$~dB/km. Then modes $A_4$ and $B_4$ turn to be modes $A_5$ and $B_5$.

\emph{Step 5}. After Charlie receives the states, he performs CV Bell detection. Specifically, modes $A_5$ and $B_5$ are interfered on a 50:50 BS, outputting modes $C$ and $D$. Subsequently, the $x-$ and $p-$quadratures of mode $C$ and mode $D$ are detected by homodyne detectors respectively, yielding $x_C$ and $p_D$. Finally, Charlie publicly announces a complex variable $\mathbf{r} =(x_C + ip_D)/\sqrt{2}$ to Alice and Bob through classical channels. After that, Alice and Bob can hence build correlations and extract secret keys after the classical postprocessing stage including parameter estimation, reconciliation and privacy amplification.

\subsubsection{Case 2: only Bob defines his prepared states}

In this case, only Bob can estimate his preparation noise due to the monitoring module, while Alice sends the states to Charlie directly. Thus, we have $T_M = 1$, $T_K =\eta_{\text M}$, and modes $F_1$, $F_2$, and $F_3$ are not involved. Similarly, the EB scheme is described as follows:

\emph{Step 1}. Same as those in \textbf{Case 1}.

\emph{Step 2}. Due to the source monitoring module, Bob is able to model his source noise by EPR states $\rho_{G_1G_2}$ so that the state Bob outputs is actually $\rho_{B_1B_3}$. Since the monitoring module is not set in Alice's side, the state that Alice outputs keeps $\rho_{A_1A_2}$. Note that if there exists RIN, then the total noise variance is $V_{\text S}+V_{\text{RIN}}$ instead of $V_S$.

\emph{Step 3}. The BS with transmittance $T_K$ in Bob separates mode $B_3$ into $B_4$ and $K_1$, while the BS with transmittance $T_M = 1$ in Alice turns mode $A_2$ to $A_4$ (note that $A_4$ is actually the same as $A_2$ in this case). In the monitoring module in Alice, subsequently, mode $K_1$ passes through the first BS with transmittance $\eta_{\text d}$, outputting modes $K_2$ and $Q_1$. Next, $K_2$ passes through the second BS with transmittance $\eta_{\text e}$, outputting modes $K_3$ and $Q_2$. Then mode $K_3$ is measured with homodyne detection.

\emph{Step 4} \& \emph{Step 5.} Same as those in \textbf{Case 1}.

\subsubsection{Case 3: both users define their prepared states}

In this case, monitoring modules are set in both sides so that Alice and Bob are both able to estimate their preparation noise. Thus, we have $T_M=T_K=\eta_{\text M}$. The EB scheme is described as:

\emph{Step 1}. Same as those in \textbf{Case 1}.

\emph{Step 2}. The states that Alice and Bob output are actually $\rho_{A_1A_3}$ and $\rho_{B_1B_3}$ with source noise modeled, respectively.

\emph{Step 3}.  The BS with transmittance $T_M$ in Alice separates mode $A_3$ into $A_4$ and $M_1$, while the BS with transmittance $T_K$ in Bob separates mode $B_3$ into $B_4$ and $K_1$. In the monitoring modules, subsequently, modes $K_1$ and $M_1$ pass through the first BS with transmittance $\eta_{\text d}$, outputting modes $K_2$ and $Q_1$, modes $M_2$ and $P_1$, respectively. Next, $K_2$ and $M_2$ pass through the second BS with transmittance $\eta_{\text e}$, outputting modes $K_3$ and $Q_2$, modes $M_3$ and $P_2$, respectively. Then modes $K_3$ and $M_3$ are measured with homodyne detection.

\emph{Step 4} \& \emph{Step 5.} Same as those in \textbf{Case 1}.

\section{Security Analysis}\label{Security analysis}

In this section, the security analysis against the collective attack in finite-size regime in three different situations are given. Without loss of generality, the secret key rate under direct reconciliation is investigated. Particularly, the secret key rate in finite-size regime conditioned on Charlie's measurement result $\mathbf{r}$ can be calculated according to Devetak-Winter theorem by~\cite{Devetak_2005_ProcRSocA_461_207}
\begin{equation}
R=\frac{n}{N}\left[\xi I_{AB|\mathbf{r}}-\left(\chi_{AE|\mathbf{r}}\right)_{\epsilon_{\text{PE}}}-\Delta(n)\right],
\end{equation}
where $N$ is the total number of signals exchanged by Alice and Bob, of which only $n$ signals are used to generate the keys. $\xi$ is the reconciliation efficiency, $I_{AB|\mathbf{r}}$ is the Shannon mutual information between Alice and Bob, and $\chi_{AE|\mathbf{r}}$ is the Holevo bound between Alice and Eve representing for the maximum amount of information Eve can acquire. Considering the influence of the finite-size effect on the accuracy of the parameter estimation, that is, under a certain failure probability ${\epsilon_{\text {PE}}}$, the realistic channel parameters are within a certain confidence interval near the estimated parameters, then the conditional entropy of Eve and Bob is expressed as $\left(\chi_{AE|\mathbf{r}}\right)_{\epsilon_{\text {PE}}}$. The most important parameter in the expression, $\Delta(n)$, is related to the security of the private amplification and is expressed as~\cite{Leverrier_2010_PhysRevA_81_062343}
\begin{equation}
\Delta(n)=\left(2 \operatorname{dim} H_{X}+3\right) \sqrt{\frac{\log _{2}(2 / \tilde{\epsilon})}{n}}+\frac{2}{n} \log _{2}\left(1 / \epsilon_{\text{PA}}\right),
\end{equation}
where the first term is the convergence rate of the smooth min-entropy of the independent identically distributed state to the von Neumann entropy, which is the main part of $\Delta(n)$. $H_X$ corresponds to the dimension of Hilbert space of variable $X$ in the raw key. Generally, $\text{dim}H_X=2$ in CV protocols. $\tilde{\epsilon}$ and $\epsilon_{\text{PA}}$ are the smoothing parameter and failure probability of the private amplification process respectively, and we take their best values as $\tilde{\epsilon}=\epsilon_{\text {PA}}=10^{-10}$~\cite{Leverrier_2010_PhysRevA_81_062343}.

Denote the result of Alice's heterodyne detection on mode $A_1$ is $\tilde{\alpha} = \left(\tilde{x}_A + i\tilde{p}_A\right)$, then the mutual information between Alice and Bob $I_{AB|\mathbf{r}}$ can be expressed as
\begin{equation}
\begin{aligned}
I_{A B \mid \mathbf{r}} &=\frac{1}{2} \log \frac{V_{B \mid \mathbf{r}}^{x}}{V_{B \mid \mathbf{r} \tilde{\alpha}}^{x}}+\frac{1}{2} \log \frac{V_{B \mid \mathbf{r}}^{p}}{V_{B \mid \mathbf{r} \tilde{\alpha}}^{p}} \\
&=\frac{1}{2} \log \left[\frac{\left(V_{B_{1} \mid \mathbf{r}}^{x}+1\right)\left(V_{B_{1} \mid \mathbf{r}}^{p}+1\right)}{\left(V_{B_{1} \mid \mathbf{r} \tilde{\alpha}}^{x}+1\right)\left(V_{B_{1} \mid \mathbf{r} \tilde{\alpha}}^{p}+1\right)}\right],
\end{aligned}
\end{equation}
where $V^{x(p)}_{B_1|\mathbf{r}}$ and $V^{x(p)}_{B_1|\mathbf{r}\tilde{\alpha}}$ are the elements on the diagonal of the covariance matrices $\gamma_{B_1|\mathbf{r}}$ and $\gamma_{B_1|\mathbf{r}\tilde{\alpha}}$ for $x(p)$-quadrature.

The Holevo bound $\chi_{AE|\mathbf{r}}$ is
\begin{equation}
\chi_{A E \mid \mathbf{r}}=S\left(\rho_{E \mid \mathbf{r}}\right)-S\left(\rho_{E \mid \mathbf{r} \tilde{\alpha}}\right),
\end{equation}
where $S \left(\rho\right)$ is the von Neumann entropy of the quantum state $\rho$. Due to the limited signal length, the statistical fluctuation of sampling estimation in the parameter estimation process will be greater, which makes the evaluation accuracy of Eve eavesdropping behavior by both communication parties worse. In order to ensure the security of the protocol, it is necessary to estimate the worst effect of eavesdropping. That is, in case of statistical fluctuation in parameter estimation, it is necessary to calculate the maximum value of Holevo information between Eve and Bob, namely, the maximum value of $\left(\chi_{AE|\mathbf{r}}\right)_{\epsilon_{\text{PE}}}$, which depends on the covariance matrix. Further, if one wants to extend the security analysis to the composable regime, just substitute the covariance matrix in the manuscript into the composable security framework \cite{Mueller-Quade_2010_NewJPhys_11_085006, Leverrier_2015_PhysRevLett_114_070501, Tomamichel_2011_NatCommun_3_634} for analysis, which will not be given more details here.

Here we consider that Eve performs the two-mode Gaussian attack in the channel, the most general attack for CV-MDI QKD (see Appendix~\ref{AppendixC} for more details). Specifically, Eve mixes her two ancillary modes $E_1$ and $E_2$ with the incident modes $A_4$ and $B_4$ through two independent BSs with transmittance $\eta_A$ and $\eta_B$ respectively, with $0\leq \eta_{A(B)}\leq1$. In the following, the secret key rate in three cases are derived in details, respectively.

\subsection{Case 1: only Alice defines her prepared states}

In this case, the monitoring module is only in Alice's side, enabling Alice to estimate her source noise. Here Bob still has no knowledge on his preparation noise due to the lack of monitoring. Therefore, modes $F_1$ and $F_3$ representing for source noise in Alice's side are supposed to be trusted, but modes $G_1$ and $G_3$ in Bob's side are not involved. For shortness, we denote $F_3F_1$ of as $F$, and $M_3P_2$ in Alice's monitoring module as $M$. Therefore, after Charlie's Bell detection, Eve is able to purify the system $\rho_{B_1A_1FME}$ and we have $S\left(\rho_{E \mid \mathbf{r}}\right)=S\left(\rho_{B_{1} A_{1} F M \mid \mathbf{r}}\right)$. After Alice's heterodyne detection with the result of $\tilde{\alpha}$, $\rho_{B_1FME}$ is a pure system thus $S\left(\rho_{E \mid \mathbf{r} \tilde{\alpha}}\right)=S\left(\rho_{B_{1} F M \mid \mathbf{r} \tilde{\alpha}}\right)$. Therefore, the Holevo bound is derived as
\begin{equation}
\chi_{A E \mid \mathbf{r}}=S\left(\rho_{B_{1} A_{1} F M \mid \mathbf{r}}\right)-S\left(\rho_{B_{1} F M \mid \mathbf{r} \tilde{\alpha}}\right),
\end{equation}
where the first and second terms on the right of the equation are calculated from the symplectic eigenvalues $\nu_{1\sim6}$ and $\nu_{7\sim11}$ of covariance matrices $\gamma_{B_1A_1FM|\mathbf{r}}$ and $\gamma_{B_1FM|\mathbf{r}\tilde{\alpha}}$. Then the Holevo bound in this case thus takes the form
\begin{equation}
\chi_{A E \mid \mathbf{r}}=\sum_{i=1}^{6} g\left(\frac{\nu_{i}-1}{2}\right)-\sum_{i=7}^{11} g\left(\frac{\nu_{i}-1}{2}\right),
\end{equation}
where $g(x)=(x+1) \log _{2}(x+1)-x \log _{2} x$.

The covariance matrix $\gamma_{B_1A_1FM|\mathbf{r}}$ is calculated by
\begin{equation}\label{Case1}
\gamma_{B_{1} A_{1} F M \mid \mathbf{r}}=\gamma_{B_{1} A_{1} F M}-\mathbf{C}^{(1)} \mathbf{R}^{(1)^{-1}} \mathbf{C}^{(1)^{\text{T}}},
\end{equation}
where $\gamma_{B_1A_1FM}$ is Alice and Bob's reduced covariance matrix, $\mathbf{R}^{(1)}$ is the covariance matrix of relay's outcomes, and $\mathbf{C}^{(1)}$ is the covariance matrix of the correlations. More details of the covariance matrices needed for the computation in three schemes are described in Appendix~\ref{AppendixD}.

Given the outcome of Alice's heterodyne detection $\tilde{\alpha}$, the conditional covariance matrix $\gamma_{B_1FM|\mathbf{r}\tilde{\alpha}}$ is derived by the expression
\begin{equation}
\begin{aligned}
\gamma_{B_{1} F M \mid \mathbf{r} \tilde{\alpha}}&=\gamma_{B_{1} F M \mid \mathbf{r}} \\
&-\sigma_{A_{1} B_{1} F M \mid \mathbf{r}}^{\text{T}}\left(\gamma_{A_{1} \mid \mathbf{r}}+\mathbf{I}\right)^{-1} \sigma_{A_{1} B_{1} F M \mid \mathbf{r}},
\end{aligned}
\end{equation}
where $\gamma_{B_1FM|\mathbf{r}}$, $\gamma_{A_1|\mathbf{r}}$ and $\sigma_{A_1B_1FM|\mathbf{r}}$ are submatrices of $\gamma_{B_1A_1FM|\mathbf{r}}$.

\begin{figure*}
	\centering
\subfigure[]{
		\label{RIN_Case1_sym}
		\includegraphics[width=0.333\linewidth]{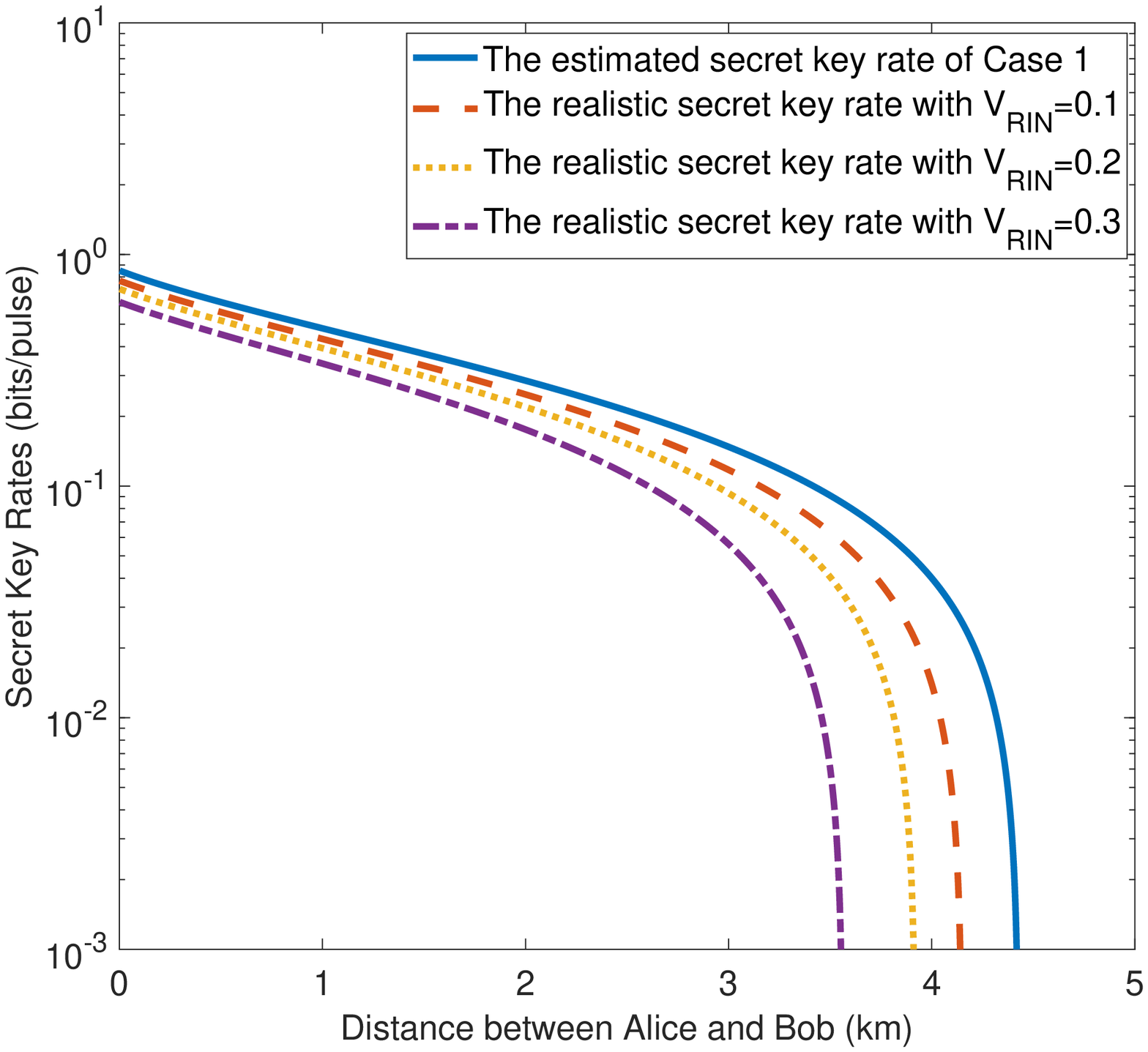}}
 \hspace{-5mm}
	\subfigure[]{
		\label{RIN_Case2_sym}
		\includegraphics[width=0.333\linewidth]{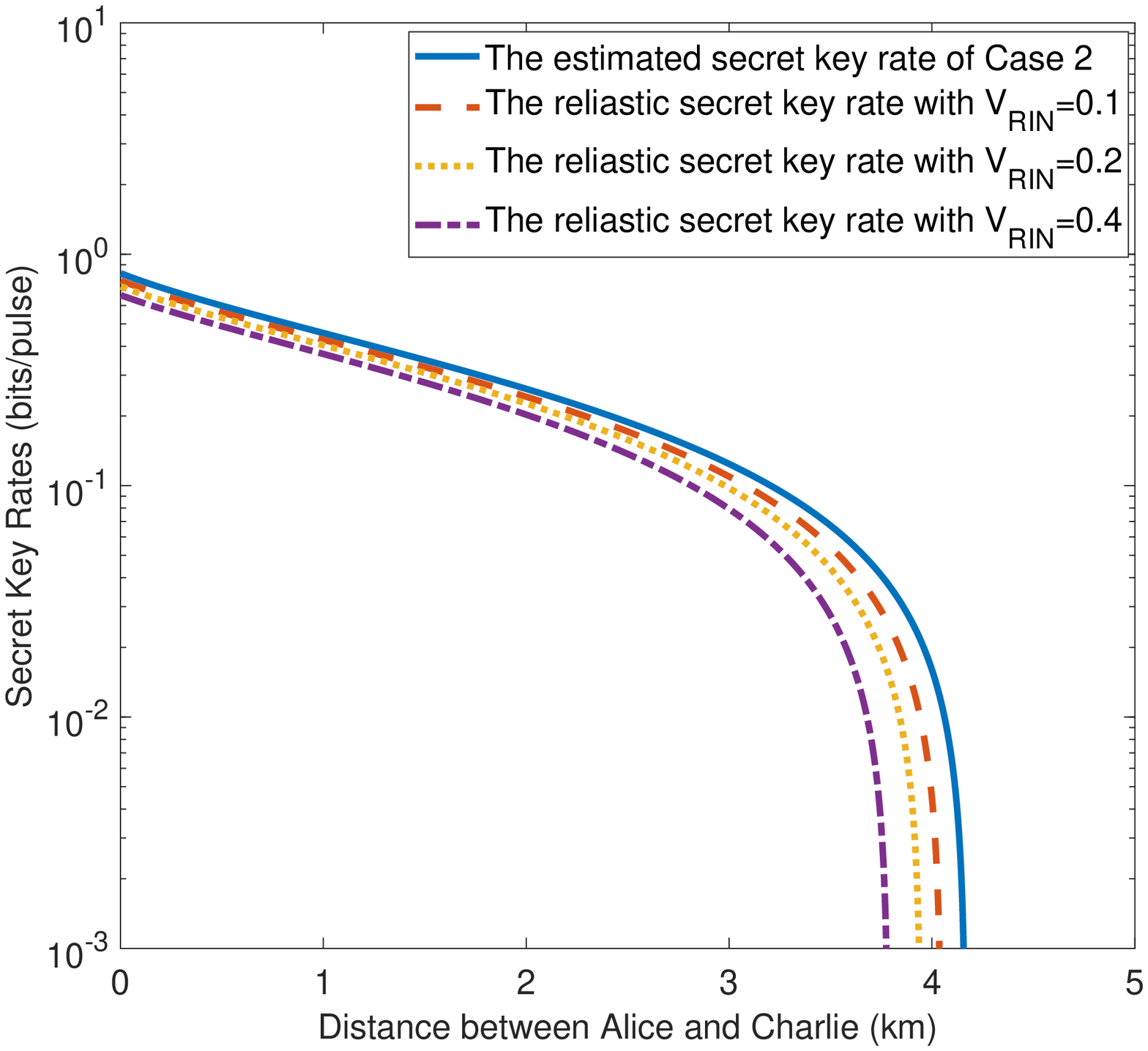}}
 \hspace{-5mm}
	\subfigure[]{
		\label{RIN_Case3_sym}
		\includegraphics[width=0.333\linewidth]{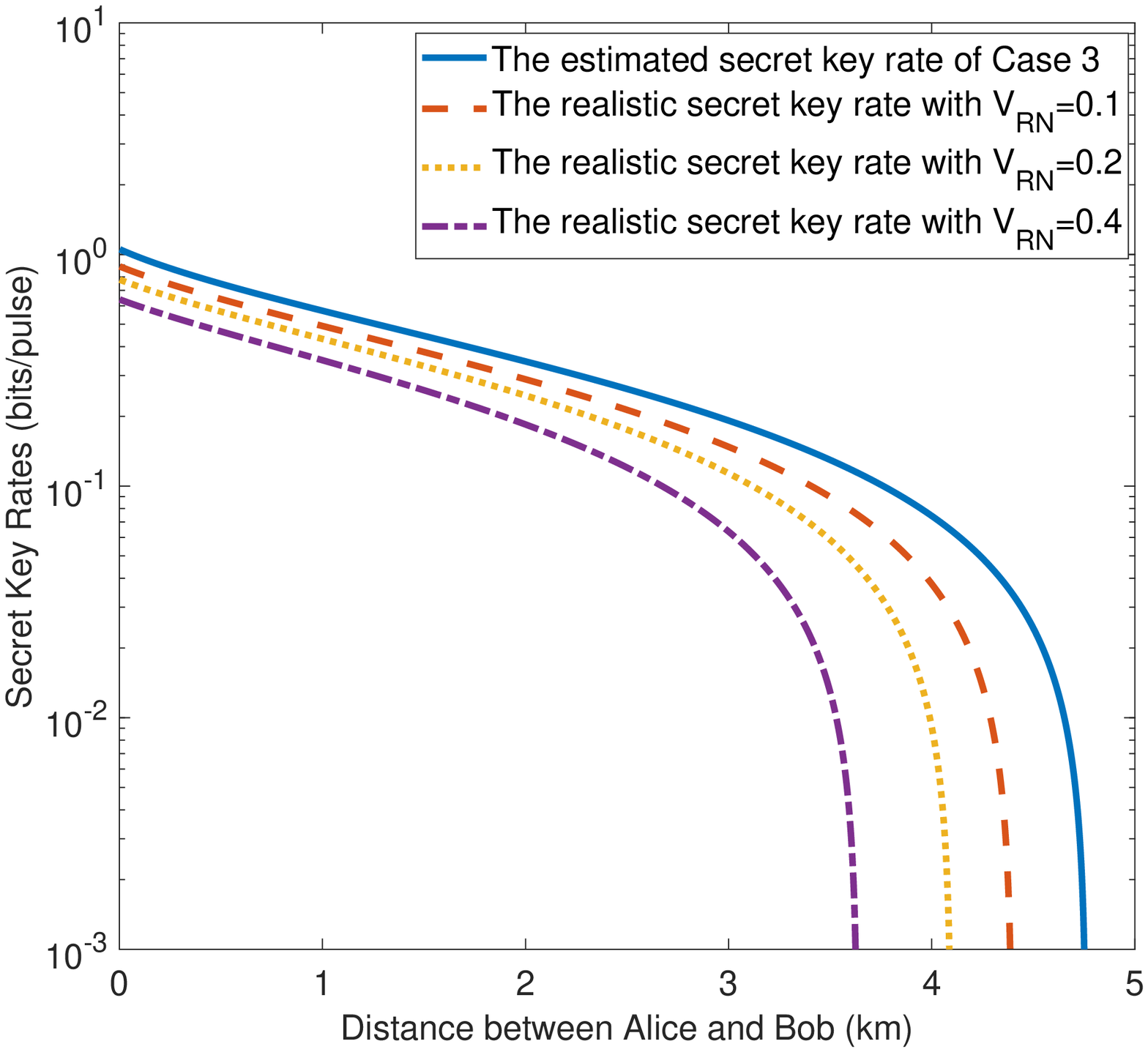}}
 \hspace{-5mm}
	\subfigure[]{
		\label{RIN_Case1_asym}
		\includegraphics[width=0.333\linewidth]{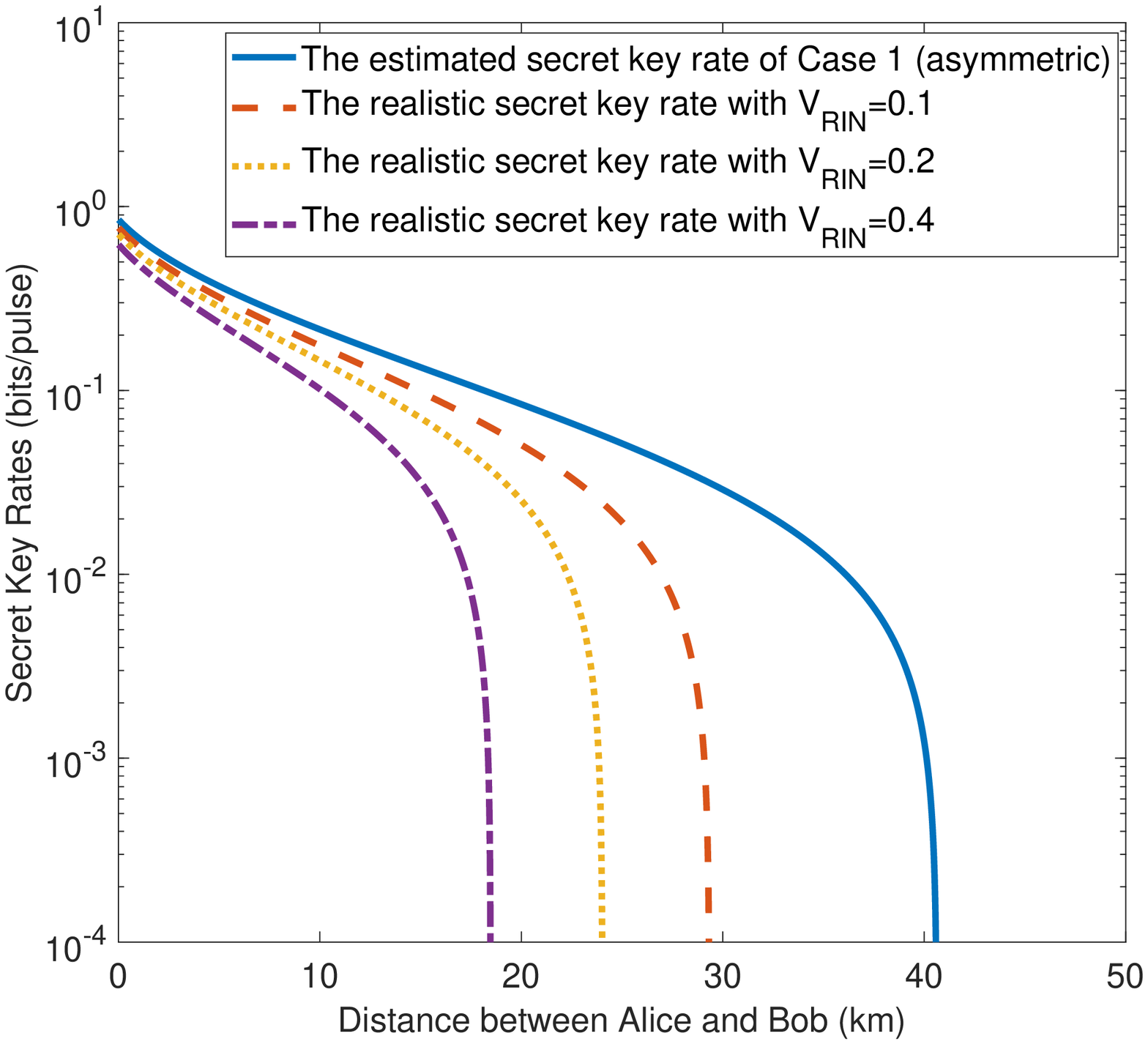}}
 \hspace{-5mm}
	\subfigure[]{
		\label{RIN_Case2_asym}
		\includegraphics[width=0.333\linewidth]{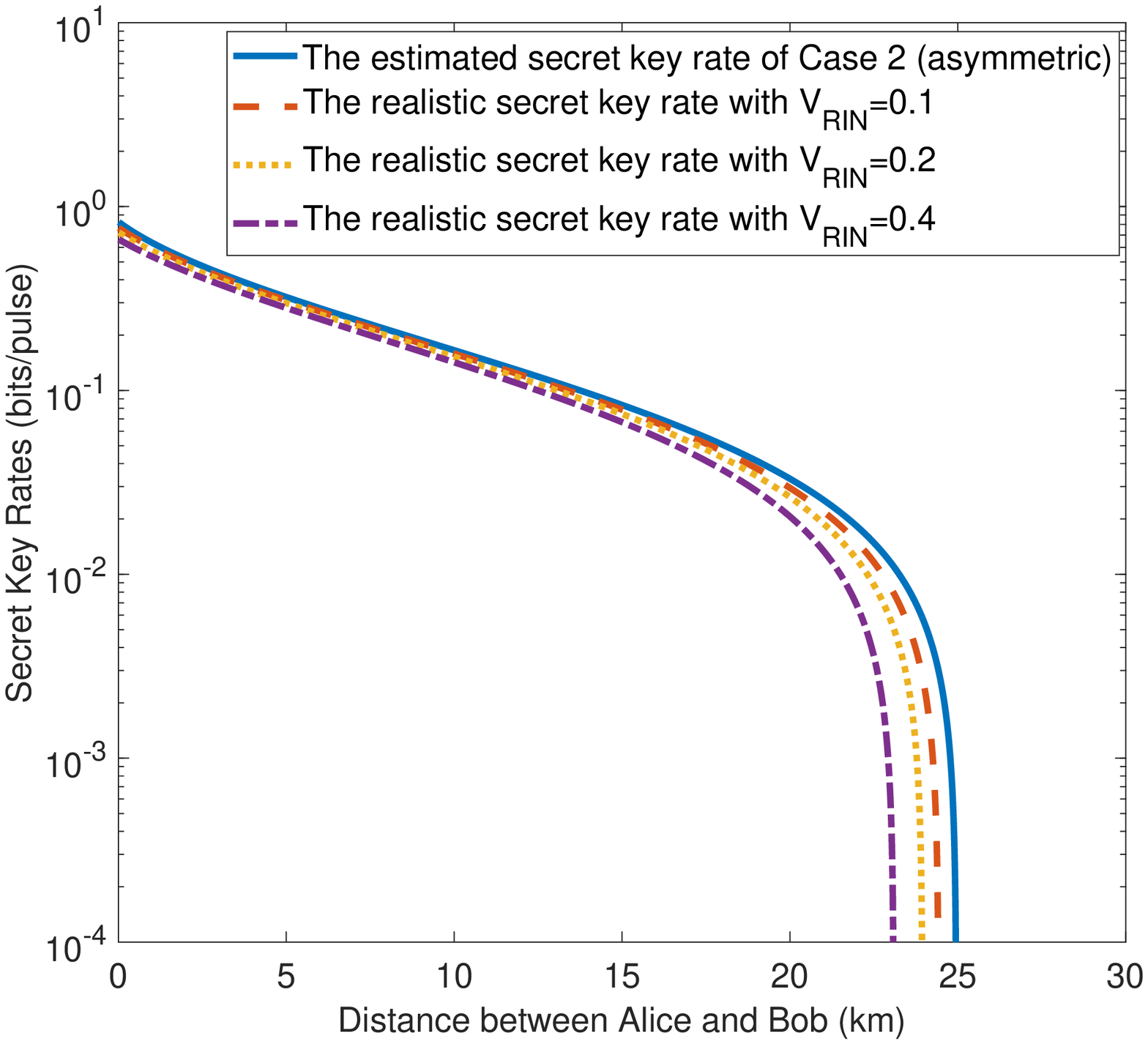}}
 \hspace{-5mm}
	\subfigure[]{
		\label{RIN_Case3_asym}
		\includegraphics[width=0.333\linewidth]{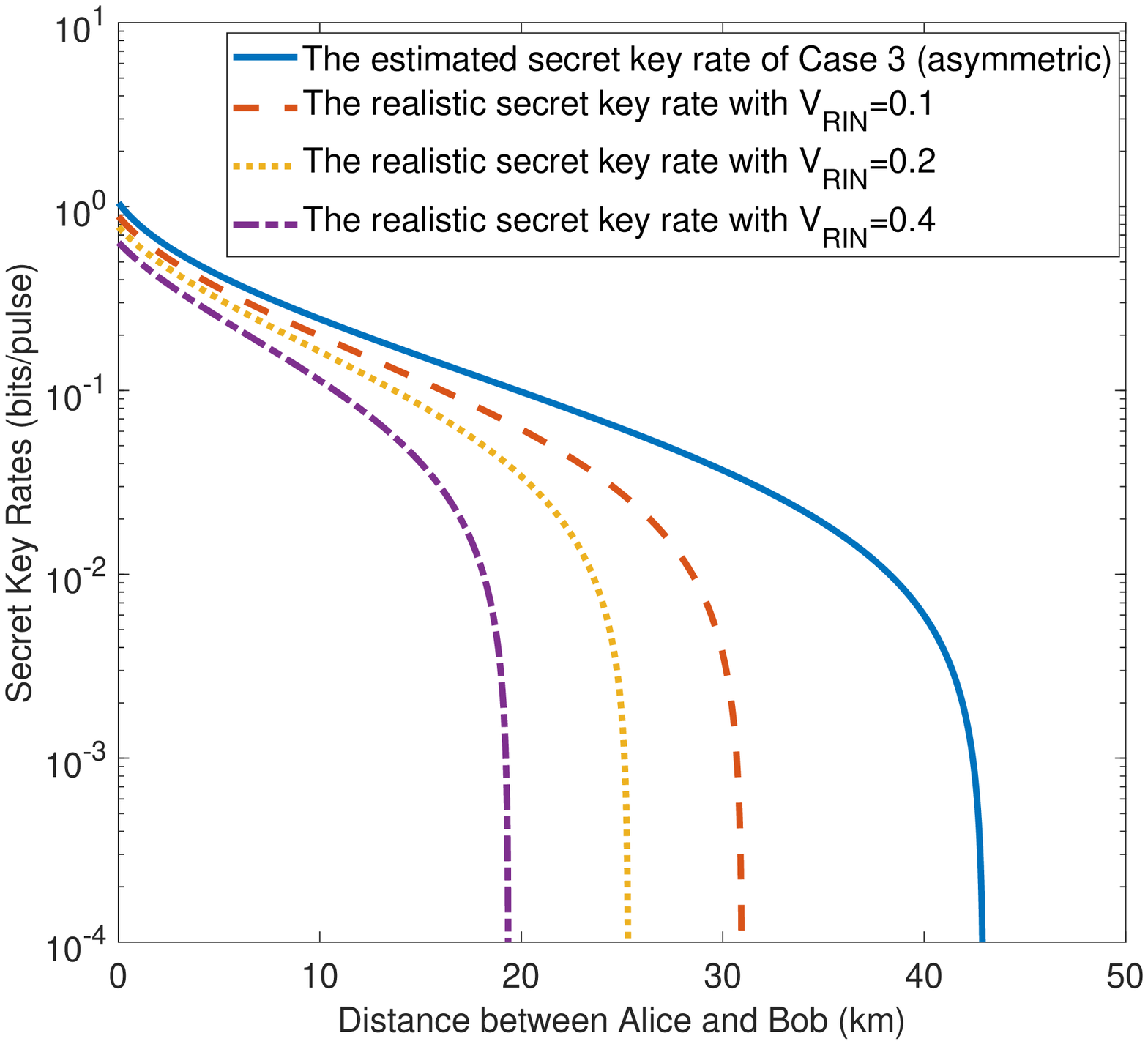}}
	\caption{Secret key rates versus distance $L_{AB}$ when (a) Alice is able to determine her preparation noise, (b) Bob is able to determine his preparation noise and (c) both users are able to determine their preparation noise in the symmetric configuration ($L_{AC}=L_{BC}$) and (d) Alice is able to determine her preparation noise, (e) Bob is able to determine his preparation noise and (f) both users are able to determine their preparation noise in asymmetric configuration ($L_{AC}\rightarrow 0$), respectively. The variance of RIN varies to be $V_{\text{RIN}}=0.1$, 0.2, and 0.4. The red dashes, yellow dotted, and purple dot-dashed lines correspond to the three cases mentioned above. In particular, each blue curve of the estimated secret key rate has the same configuration and monitoring module settings as other lines in each subfigure, but RIN is ignored. The reconciliation efficiency $\xi=1$, modulation variance $V_{\text{Mod}}= 60$, excess noise $\varepsilon= 0.01$, transmittance and variance of EPR state in the source noise model $T_{\text S} = 0.99$, $V_{\text S} = 3$. In the monitoring module, the transmittance of the BS $\eta_{\text M} = 0.9$, detection efficiency $\eta_{\text{d}} = 0.6$, electronic noise $v_{\text{el}} = 0.01$. Here the finite-size effect is considered and the block length is fixed as $10^8$.}
	\label{RIN}
\end{figure*}

\subsection{Case 2: only Bob defines his prepared states}

Similarly, knowing what is practically prepared enables Bob to know the source noise while Alice still has no knowledge on her source noise due to the lack of monitoring. Therefore, modes $G_1$ and $G_3$ representing for source noise are supposed to be trusted, while modes $F_1$ and $F_3$ are not involved. For shortness, we denote $G_3G_1$ of as $G$, and $K_3Q_2$ in Bob's monitoring module as $K$. Therefore, after Charlie's Bell detection, Eve is able to purify the system $\rho_{B_1A_1GKE}$ and we have $S\left(\rho_{E \mid \mathbf{r}}\right)=S\left(\rho_{B_{1} A_{1} GK \mid \mathbf{r}}\right)$. After Alice's heterodyne detection, $\rho_{B_1GKE}$ is a pure system thus $S\left(\rho_{E \mid \mathbf{r} \tilde{\alpha}}\right)=S\left(\rho_{B_{1} GK \mid \mathbf{r} \tilde{\alpha}}\right)$. Therefore, the Holevo bound is derived as
\begin{equation}
\chi_{A E \mid \mathbf{r}}=S\left(\rho_{B_{1} A_{1} GK \mid \mathbf{r}}\right)-S\left(\rho_{B_{1} GK \mid \mathbf{r} \tilde{\alpha}}\right),
\end{equation}
where $S\left(\rho_{B_{1} A_{1} GK \mid \mathbf{r}}\right)$ and $S\left(\rho_{B_{1} GK \mid \mathbf{r} \tilde{\alpha}}\right)$ are calculated from the symplectic eigenvalues $\nu_{1\sim6}$ and $\nu_{7\sim11}$ of covariance matrices $\gamma_{B_1A_1GK|\mathbf{r}}$ and $\gamma_{B_1GK|\mathbf{r}\tilde{\alpha}}$. Then the Holevo bound in this case thus takes the form
\begin{equation}
\chi_{A E \mid \mathbf{r}}=\sum_{i=1}^{6} g\left(\frac{\nu_{i}-1}{2}\right)-\sum_{i=7}^{11} g\left(\frac{\nu_{i}-1}{2}\right).
\end{equation}

The covariance matrix $\gamma_{B_1A_1GK|\mathbf{r}}$ conditioned on Charlie's measurement results $\mathbf{r}$ is calculated by
\begin{equation}\label{Case2}
\gamma_{B_{1} A_{1} GK \mid \mathbf{r}}=\gamma_{B_{1} A_{1} GK}-\mathbf{C}^{(2)} \mathbf{R}^{(2)^{-1}} \mathbf{C}^{(2)^{\text{T}}},
\end{equation}
where $\gamma_{B_1A_1GK}$ is Alice and Bob's reduced covariance matrix, $\mathbf{R}^{(2)}$ is the covariance matrix of relay's outcomes, and $\mathbf{C}^{(2)}$ is the covariance matrix of the correlations.

After Alice heterodyne detects her remained mode, the conditional covariance matrix $\gamma_{B_1GK|\mathbf{r}\tilde{\alpha}}$ is derived by
\begin{equation}
\begin{aligned}
\gamma_{B_{1} GK \mid \mathbf{r} \tilde{\alpha}}&=\gamma_{B_{1} GK \mid \mathbf{r}} \\
&-\sigma_{A_{1} B_{1} GK \mid \mathbf{r}}^{\text{T}}\left(\gamma_{A_{1} \mid \mathbf{r}}+\mathbf{I}\right)^{-1} \sigma_{A_{1} B_{1} GK\mid \mathbf{r}},
\end{aligned}
\end{equation}
where $\gamma_{B_1GK|\mathbf{r}}$, $\gamma_{A_1|\mathbf{r}}$ and $\sigma_{A_1B_1GK|\mathbf{r}}$ are submatrices of $\gamma_{B_1A_1GK|\mathbf{r}}$.

\subsection{Case 3: both users define their prepared states}

In this case, Alice and Bob both utilize monitoring modules which enable them to know the source noise. Therefore, modes $F_1$, $F_3$, $G_1$ and $G_3$ representing for source noise are supposed to be trusted. Therefore, after Charlie's Bell detection, Eve is able to purify the system $\rho_{A_1FMB_1GKE}$ hence $S\left(\rho_{E \mid \mathbf{r}}\right)=S\left(\rho_{A_1FMB_1GK\mid \mathbf{r}}\right)$. After Alice's heterodyne detection with the result of $\tilde{\alpha}$, $\rho_{FMB_1GKE}$ is a pure system thus $S\left(\rho_{E \mid \mathbf{r} \tilde{\alpha}}\right)=S\left(\rho_{FMB_{1} GK \mid \mathbf{r} \tilde{\alpha}}\right)$. Thus,
\begin{equation}
\chi_{A E \mid \mathbf{r}}=S\left(\rho_{A_{1}FMB_1 GK \mid \mathbf{r}}\right)-S\left(\rho_{FMB_{1} GK \mid \mathbf{r} \tilde{\alpha}}\right),
\end{equation}
where $S\left(\rho_{A_{1}FMB_1 GK \mid \mathbf{r}}\right)$ and $S\left(\rho_{FMB_{1} GK \mid \mathbf{r} \tilde{\alpha}}\right)$ are calculated from the symplectic eigenvalues $\nu_{1\sim10}$ and $\nu_{11\sim19}$ of covariance matrices $\gamma_{A_1FMB_1GK|\mathbf{r}}$ and $\gamma_{FMB_1GK|\mathbf{r}\tilde{\alpha}}$. Then the Holevo bound in this case thus takes the form
\begin{equation}
\chi_{A E \mid \mathbf{r}}=\sum_{i=1}^{10} g\left(\frac{\nu_{i}-1}{2}\right)-\sum_{i=11}^{19 } g\left(\frac{\nu_{i}-1}{2}\right).
\end{equation}

Similarly, the covariance matrix $\gamma_{A_1FMB_1GK|\mathbf{r}}$ conditioned on Charlie's measurement results $\mathbf{r}$ is calculated by
\begin{equation}\label{Case3}
\gamma_{A_{1}FMB_1 GK \mid \mathbf{r}}=\gamma_{A_{1}FMB_1 GK}-\mathbf{C}^{(3)} \mathbf{R}^{(3)^{-1}} \mathbf{C}^{(3)^{\text{T}}},
\end{equation}
where $\gamma_{A_1FMB_1GK}$ is Alice and Bob's reduced covariance matrix, $\mathbf{R}^{(3)}$ is the covariance matrix of relay's outcomes, and $\mathbf{C}^{(3)}$ is the covariance matrix of the correlations.

Further, the conditional covariance matrix $\gamma_{FMB_1GK|\mathbf{r}\tilde{\alpha}}$ is derived by
\begin{equation}
\begin{aligned}
\gamma_{FMB_{1} GK \mid \mathbf{r} \tilde{\alpha}}&=\gamma_{FMB_{1} GK \mid \mathbf{r}} \\
&-\sigma_{A_{1}FM B_{1} GK \mid \mathbf{r}}^{\text{T}}\left(\gamma_{A_{1} \mid \mathbf{r}}+\mathbf{I}\right)^{-1} \sigma_{A_{1} FMB_{1} GK\mid \mathbf{r}},
\end{aligned}
\end{equation}
where $\gamma_{FMB_1GK|\mathbf{r}}$, $\gamma_{A_1|\mathbf{r}}$ and $\sigma_{A_1FMB_1GK|\mathbf{r}}$ are submatrices of $\gamma_{A_1FMB_1GK|\mathbf{r}}$.

Notice that in the practical CV-MDI QKD protocol, we need to consider the statistical fluctuation of the channel transmittance $\left(\eta_A, \eta_B\right)$, the excess noise $\varepsilon_1$ of the Alice-Charlie channel and $\varepsilon_2$ of the Bob-Charlie channel, the transmittance $T_{\text S}$, $T_M$, $T_K$, and the variance of source noise $V_{\text S}=1+T_{\text S}\varepsilon_{\text S}/\left(1-T_{\text S}\right)$. These parameters can be estimated in the parameter estimation step, which is revealed in details in Appendix~\ref{AppendixE}.

\section{Results and Discussion}\label{Results}

In this section, the performance of the aforementioned three cases for CV-MDI QKD are analyzed according to numerical simulation. Specifically, the secret key rates with finite-size effect corresponding to three cases are plotted as a function of the transmission distance from Alice to Bob $L_{AB}=L_{AC}+L_{BC}$ against the general two-mode attack. In particular, the secret key rate representing for the case with untrusted source noise is also plotted for comparison. That is, the noise is not modeled to be trusted and totally controlled by Eve. The transmittance $\eta_{\text M}=0.9$, reconciliation efficiency $\xi = 1$, modulation variance $V_{\text{Mod}} = 60$, and excess noise $\varepsilon = 0.01$. The transmittance and variance of EPR state in the source noise model are $T_{\text S} = 0.99$ and $V_{\text S} = 3$. Assume that the practical detection efficiency $\eta_{\text d} = 0.6$ and electronic noise $v_{\text{el}} = 0.01$ in the monitoring module. In all simulation, the block length is fixed as $10^8$.

\begin{figure}
\includegraphics[width=3.5in]{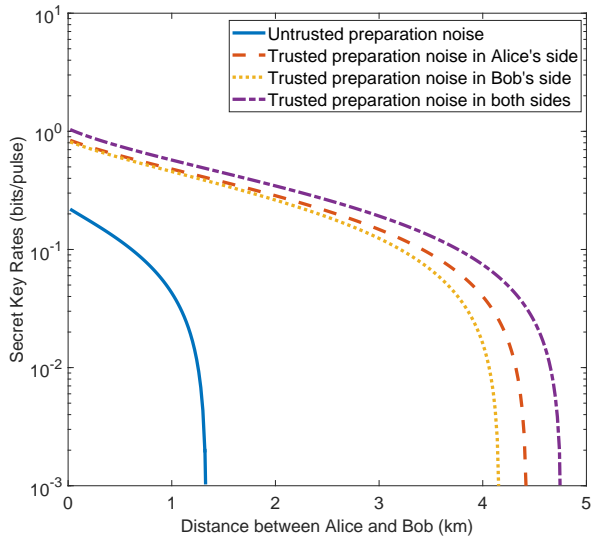}
\caption{Secret key rates versus the transmission distance in the aforementioned three cases when the distance between Alice and Charlie is equal to the distance between Bob and Charlie, that is, $L_{AC}=L_{BC}$. The simulation parameters are set as before. Particularly, the case where the source noise is untrusted is also illustrated for comparison, corresponding to the blue solid line. }\label{Rateviadistace_sym}
\end{figure}

Here the impact of RIN is investigated first to show the differences between the cases that RIN is considered or not. As shown in Fig.~\ref{RIN}, the blue solid lines represent for the case that RIN exists but is not considered (see Appendix~\ref{AppendixA} for details), while the red dashed lines, the yellow dotted lines and purple dot-dashed lines represent for $V_{\text{RIN}}=0.1, 0.2$ and 0.4, respectively. The Figs.~\ref{RIN_Case1_sym},~\ref{RIN_Case2_sym} and~\ref{RIN_Case3_sym} separately correspond to that only Alice is able to determine her source noise, only Bob is able to determine his source noise, and both users are able to determine their source noise, respectively, in symmetric configuration $L_{AC}=L_{BC}$. Here determining the source noise with the monitoring module means that due to the monitoring, the noise is under control of the users, thus they can conduct a trusted model of it, as shown in Fig.~\ref{PM&EBscheme}(b). Figs.~\ref{RIN_Case1_asym}, \ref{RIN_Case2_asym} and \ref{RIN_Case3_asym} correspond to the three cases mentioned above in asymmetric configuration, in which Charlie is set near one of the senders. Without loss of generality, we assume $L_{AC} \rightarrow 0$, then we have $\eta_A \rightarrow 1$, while $\eta_B=10^{-\alpha L_{BC}/10}$. According to simulation results, firstly, the blue curves of the three cases in the symmetric and asymmetric configurations are different. In fact, the three blue curves in the symmetric configuration correspond to the respective cases that Alice sets the monitoring module, Bob sets the monitoring module and both the users set the monitoring module. The three blue curves in the asymmetric configuration also correspond to the cases mentioned above. Due to different settings of monitoring module, the final covariance matrices are different. Thus, the blue curves perform different in subfigures. Secondly, the results show that if Alice and Bob ignore the existence of the RIN, they will overestimate the secret key rate in all cases no matter it is symmetric configuration or asymmetric configuration. That is to say, ignoring the RIN will result in the users overestimating the secret key rates and leaving a security loophole. At this time, the eavesdropper can utilize it to attain secure information quietly.

Here we take the asymmetric configuration for an example. In the case that only Alice can estimate her preparation noise, the estimated secret key rate are about 1.5 times, 2.5 times, and 26.6 times higher than the realistic rates at 18 km with $V_{\text{RIN}}=0.1, 0.2$ and 0.4, respectively. In the case that only Bob can define his prepared states, the estimated secret key rate are about 1.1 times, 1.2 times, and 1.4 times higher than the realistic rates at 18 km with $V_{\text{RIN}}=0.1, 0.2$ and 0.4, respectively. In the case that both users can determine their preparation noise, the estimated secret key rate are about 1.5 times, 2.3 times, and 10.7 times higher than the realistic rates at 18 km with $V_{\text{RIN}}=0.1, 0.2$ and 0.4, respectively. In other words, as the variance of RIN increases, the difference between the practical secret key rate and the estimated secret key rate increases. Namely, as the variance of RIN increases, the secure information that Eve are able to obtain increases. Thus, the larger the RIN, the greater the security vulnerability of the CV-MDI QKD system caused by ignoring the RIN. Fortunately, using the proposed scheme is beneficial to avoid this underlying threat because the RIN is adopted to be part of the SNU.  Besides, compared Figs.~\ref{RIN_Case2_sym} and \ref{RIN_Case2_asym} with Figs.~\ref{RIN_Case1_sym} and \ref{RIN_Case1_asym}, it is seen that the system performs better in the case where the monitoring module is utilized only in Alice’s side than in the case where monitoring module is utilized only in Bob’s side. It is due to the "fighting noise with noise" effect~\cite{GarcaPatron_2009_PhysRevLett_102_130501}, and this phenomenon only holds when the RIN is of comparatively small value. In this situation, the optimal performance is achieved when both users are able to estimate their preparation noise.

\begin{figure}
\includegraphics[width=3.5in]{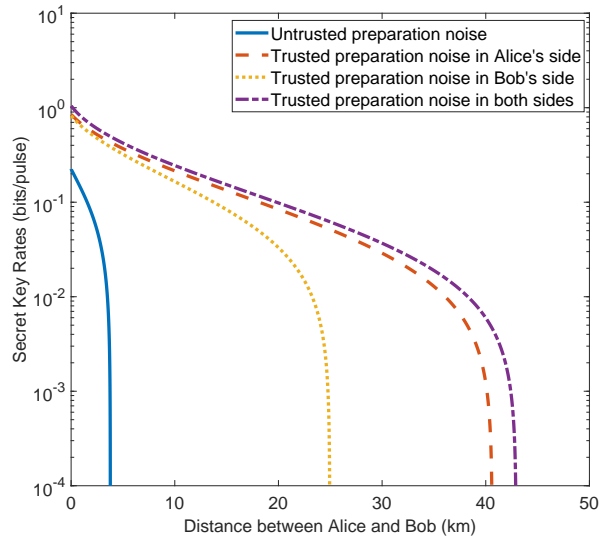}
\caption{Secret key rates versus the transmission distance in the aforementioned three cases when Charlie is put near Alice, that is, $L_{AC}\rightarrow 0$. Similarly, the red dashed, yellow dotted,  purple dot-dashed lines correspond to the three cases where only Alice can define her preparation noise, only Bob can define his preparation noise, and both users can define their preparation noise. Particularly, the case where the source noise is untrusted is also illustrated for comparison, corresponding to the blue solid line. The other parameters are set as same as the symmetric condition.}\label{Rateviadistace_asym}
\end{figure}

Next, the comparison among the aforementioned three cases and the case where the source noise is untrusted is given. At this time, the monitoring module is not added, thus the preparation noise is unable to be modeled, and the covariance matrix between Alice and Bob after Charlie's measurement is just $\gamma_{A_1B_1|\mathbf{r}}$. Here we set $V_{\text{RIN}}=0$ to show the pure advantage of trusted modeling using our proposed monitoring module. As shown in Fig.~\ref{Rateviadistace_sym}, the blue solid line, the red dashed line, yellow dotted line, and purple dot-dashed line correspond to the four cases where there the source noise is undefined without monitoring module of our proposed scheme, monitoring module is utilized in Alice's side, Bob's side and both sides in symmetric configuration, respectively. The parameters in the simulation are set as mentioned before. It is apparent that the performance with monitoring is improved, and the optimal performance is achieved in \textbf{Case 3} where Alice and Bob both implement monitoring. Subsequently, the asymmetric condition, which attains a better performance, is considered. The simulation results are depicted in Fig.~\ref{Rateviadistace_asym}. According to the results, a similar conclusion is attained that adding the monitoring module in the system helps to model the source noise and improve the performance. Besides, when Alice and Bob can both define their preparation noise, the improvement achieves optimal. Intuitively, this is because more source noises are established to trusted noises in this case.

\begin{figure*}
	\centering
\subfigure[]{
		\label{etaM_Case1}
		\includegraphics[width=0.33\linewidth]{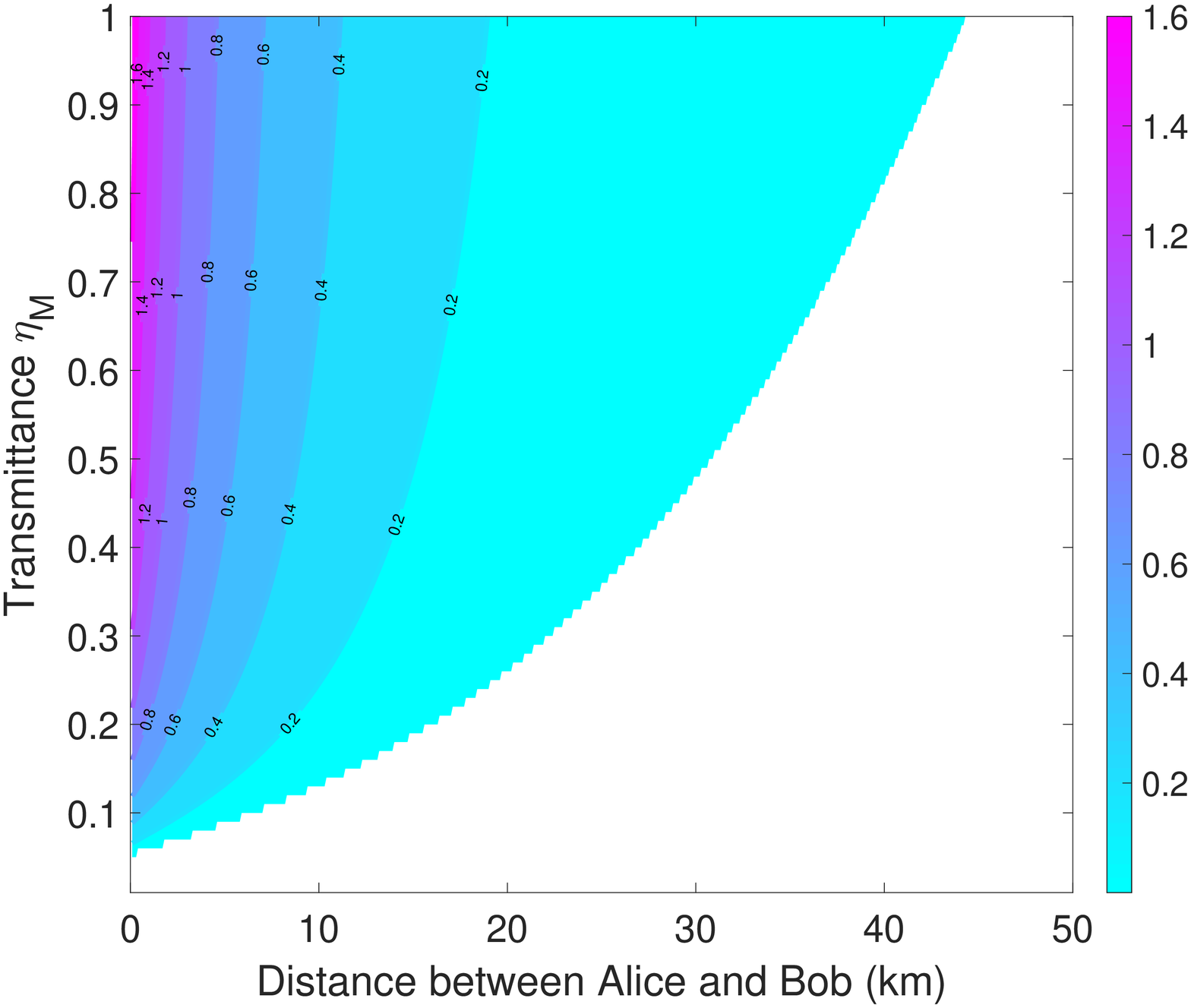}}
 \hspace{-5mm}
	\subfigure[]{
		\label{etaM_Case2}
		\includegraphics[width=0.33\linewidth]{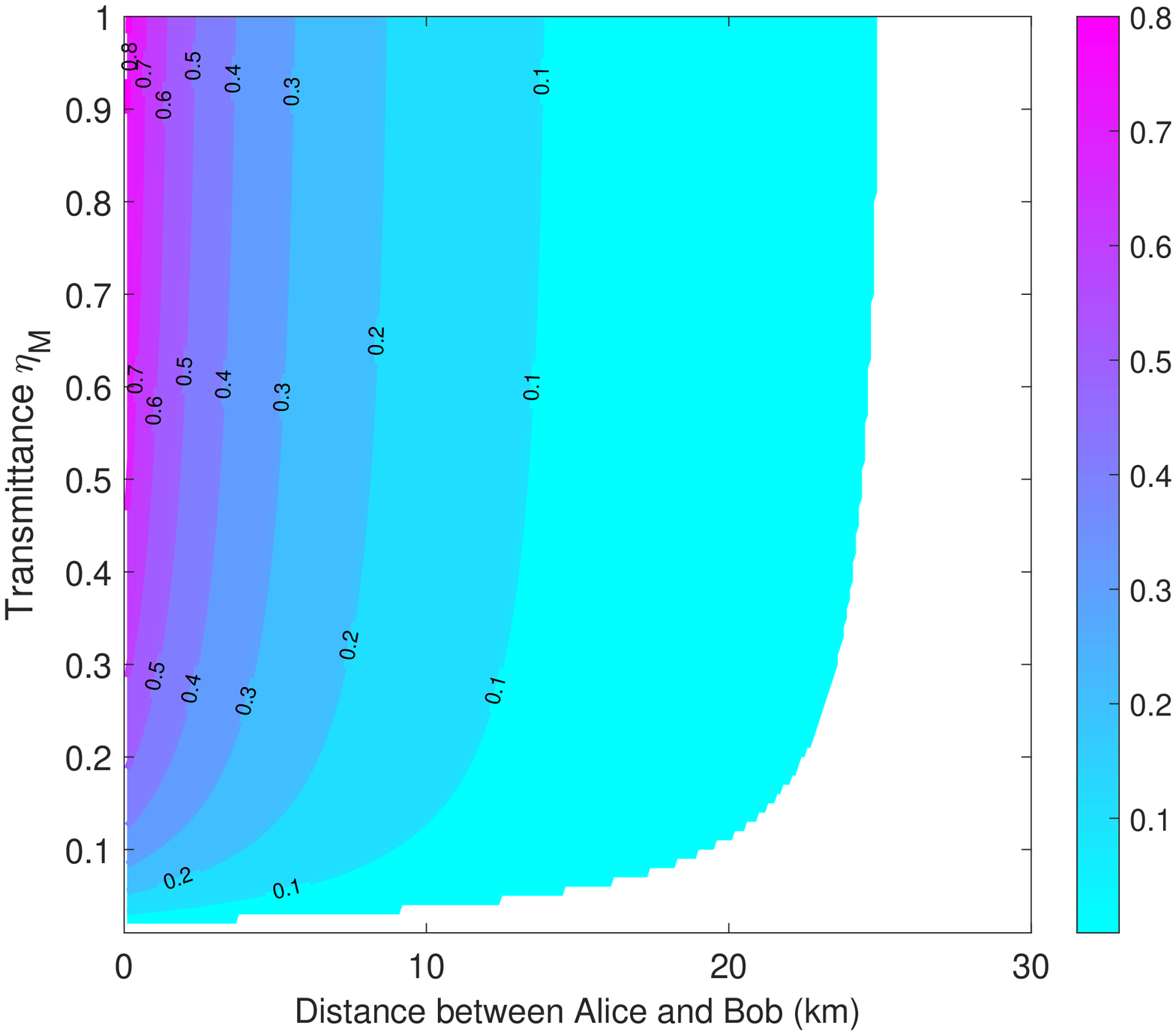}}
 \hspace{-5mm}
	\subfigure[]{
		\label{etaM_Case3}
		\includegraphics[width=0.33\linewidth]{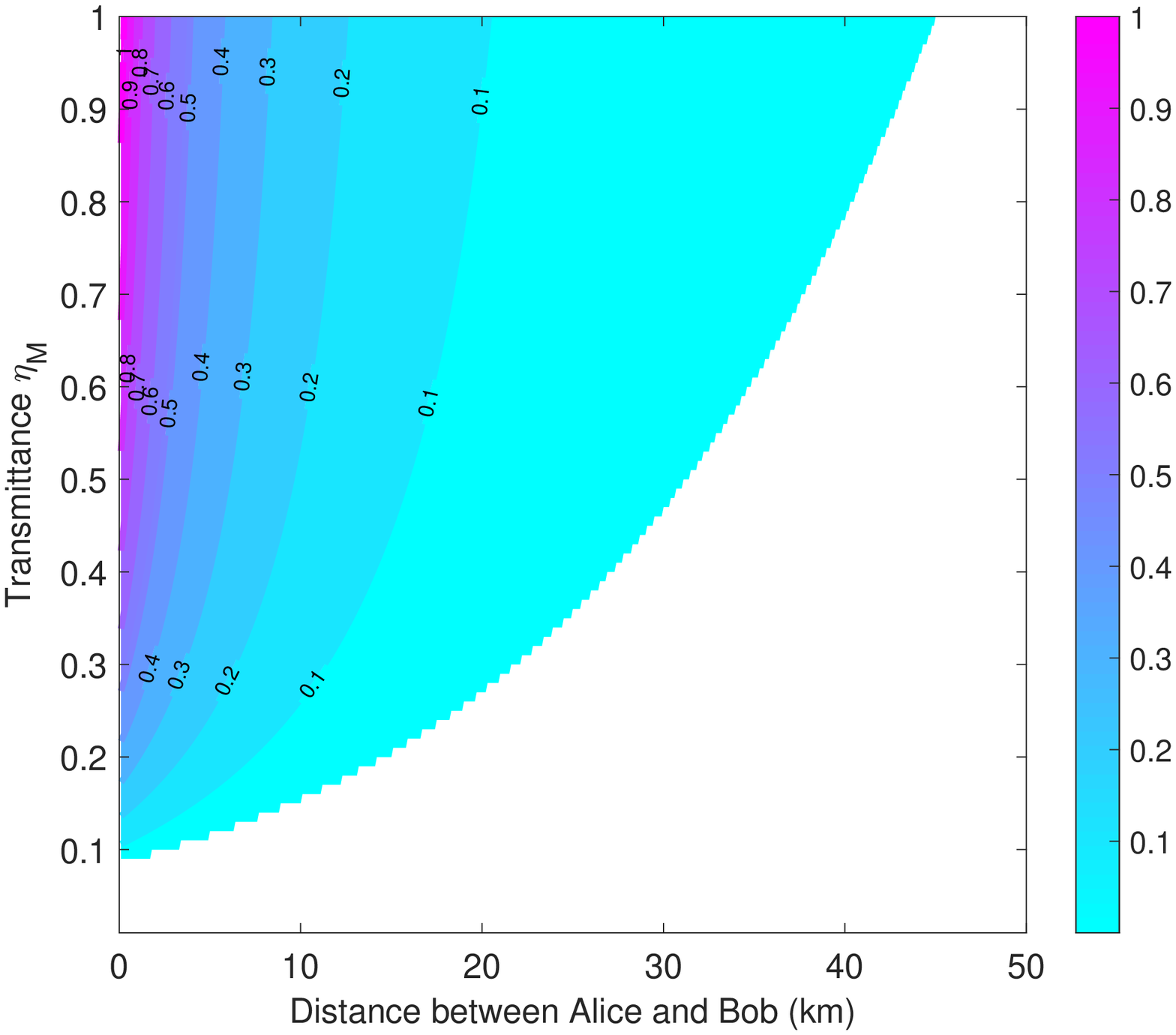}}
	\caption{Secret key rate versus distance $L_{AB}$ and transmittance $\eta_{\text M}$ when (a) Alice can define her preparation noise, (b) Bob can define his preparation noise and (c) both sides can define their preparation noise in the asymmetric situation, that is, $L_{AC}\rightarrow 0$. In particular, the condition $V_{\text{RIN}}=0$ is considered to show the pure impact of transmittance $\eta_{\text M}$ of the monitoring module. The other parameters are set as same as before.}
	\label{etaM}
\end{figure*}

To further improve the performance of the proposed scheme, we investigate the effect of the BS transmittance $\eta_{\text M}$ in the monitoring module. Here we also consider the asymmetric configuration. Fig.~\ref{etaM} illustrates the secret key rates as a function of the distance $L_{AB}$ and transmittance $\eta_{\text M}$ corresponding to the aforementioned three cases in the asymmetric configurations. From the results we can see that the optimal selection of the transmittances $\eta_{\text M}$ varies with the place where the monitoring is employed. As shown in Fig.~\ref{etaM}, the highest secret key rate and longest secure transmission distance are achieved when $\eta_{\text M}$ approach 1. In addition, when $\eta_{\text M}$ is fixed, the results of \textbf{Case 1} and \textbf{Case 3} are better than that of \textbf{Case 2}. In \textbf{Case 1} and \textbf{Case 3}, the maximal transmission distances are above 42.7 km and 45.1 km respectively, achieved when $\eta_{\text M}$ is close to 1. Things are slightly different when employing monitoring only in Bob's side, for the maximal secure distance is 24.9 km in the same condition. However, the change of $\eta_{\text M}$ does not have obvious impact as in \textbf{Case 1} and \textbf{Case 3}. In \textbf{Case 2}, the transmission distance can still achieve 24.4 km when $\eta_{\text M}$ becomes 0.5, remaining almost constant compared with the maximal distance. Even when $\eta_{\text M}$ becomes 0.2, the transmission distance can still achieve 22.4 km, decreasing by 2 km only.

\section{Conclusion}\label{Conclusion}

In this paper, we propose a countermeasure for negative impacts induced from the relative intensity noise in the continuous-variable measurement-device-independent quantum key distribution protocol based on one-time shot-noise-unit calibration method. Particularly, according to the vulnerability of the practical sources (easy or not to be attacked), three cases where only Alice is able to determine her preparation noise, only Bob is able to determine his preparation noise as well as both users are able to determine their preparation noise, are investigated in details respectively. The proposed scheme involves the relative intensity noise of the laser, eliminating the potential security loopholes. The simulation results show that the estimated secret key rate without our proposed scheme, corresponding to ignoring the relative intensity noise, is about 10.7 times higher than the realistic rate, corresponding to Alice and Bob both estimate their source noise with our proposed scheme, at 18 km transmission distance when the variance of RIN is only 0.4. What's worse, the difference becomes greater and greater with the increase of the variance of RIN. Thus, our proposed scheme makes sense in further completing the practical security of CV-MDI QKD system. Further, the relation between the secret key rate and the beam splitter transmittance of monitoring is investigated, indicating the condition for optimizing the schemes efficiently. Thus, our work enables CV-MDI QKD system not only to resist all attacks against detectors, but also to close the vulnerability caused by the actual source, thus making the scheme closer to practical security.

\begin{acknowledgements}
This work was supported by the Key Program of National Natural Science Foundation of China under Grant
No. 61531003, National Natural Science Foundation of China under Grants No. 62001041 and No. 62201012,   the Fundamental Research Funds of BUPT under Grant No. 2022RC08, the Fund of State Key Laboratory of Information Photonics and Optical Communications under Grant No. IPOC2022ZT09, and Beijing University of Posts and Telecommunications Excellent Ph. D. Students Foundation under Grant No. CX2021138.
\end{acknowledgements}

\begin{appendix}
\section{Threats of neglecting RIN in source}\label{AppendixA}

The practical state the sender really prepares is acknowledged by monitoring module using the homodyne detector, which has two inherent imperfections in practice, namely the finite detection efficiency and electronic noise. In the original EB version, the two imperfections are modeled by a BS with transmittance $\eta_{\text d}$ and an EPR state with variance $v_{\text N} = 1 + v_{\text{el}}/ (1 - \eta_{\text d})$, respectively. In this model, one mode of the EPR state is coupled with the monitoring signal through the BS. Correspondingly, the homodyne detector output in the P\&M version thus would be written as
\begin{equation}\label{appendixequ:1}
X_{\text{ori}}=A X_{\text{LO}}\left(\sqrt{\eta_{\text{d}}} \hat{x}_{M}+\sqrt{1-\eta_{\text{d}}} \hat{x}_{{v_1}}\right)+X_{\text{el}},
\end{equation}
where $A$ is amplification coefficient, $X_{\text{LO}}$ and $X_{\text{el}}$ are variables of the LO and electronic noise respectively. Here $\hat{x}_{M}$ is the mode input into the monitoring detector, and $\hat{x}_{{v_1}}$ is a vacuum state. Then the result is normalized by the SNU $u = A^2X^2_{\text{LO}}$,
\begin{equation}
x_{\text{ori}}=\frac{X_{\text{ori}}}{\sqrt{u}}=\left(\sqrt{\eta_{\text{d}}} \hat{x}_{M}+\sqrt{1-\eta_{\text{d}}} \hat{x}_{{v_1}}\right)+\frac{X_{\text{el}}}{A X_{\text{LO}}}.
\end{equation}

In the practical implementation, the SNU calibration in the original method includes two steps. In details, the variance of homodyne detector output when the LO and signal path are both switched off is measured firstly. The result is denoted as the electronic noise variance $V_{\text{el}}$ without normalization. Secondly, the variance of homodyne detector output with only LO switched on is obtained. The result is denoted as the so-called total noise variance $V_{\text{tot}}$ without normalization, which is considered to contain the shot-noise and the electronic noise. It is straightforward to know the estimated SNU is $\bar{u} = V_{\text{tot}}-V_{\text{el}}$. However, the existence of RIN due to power fluctuation of the LO thus results in the change of realistic SNU. Thus, to involve the impact of RIN is significant. Considering the RIN, the output of the homodyne detector in the second step should have been $V_{\text{tot}} = u + V_{\text{el}} + V_{\text{RIN}}$. Here we have $u^{\prime} = u + V_{\text{RIN}}$, that is to say, the extra RIN has been involved. In this way, the equation (\ref{appendixequ:1}) has become

\begin{equation}
X_{\text{ori}}^{\prime}\!=\!A X_{\text{LO}}\!\left(\!\sqrt{\eta_{\text d}} \hat{x}_{M}\!+\!\sqrt{1\!-\!\eta_{\text d}} \hat{x}_{{v_1}}\!\right)\!\!+\!X_{\text{el}}\!+\!X_{\text{RIN}},
\end{equation}
where $X_{\text{RIN}}$ represents for the variable of the RIN. This output after normalized by $u^{\prime}$ can be derived as

\begin{equation}
\begin{split}\label{appendixequ:2}
x_{\text{ori}}^{\prime}=\frac{X_{\text{ori}}^{\prime}}{\sqrt{u^{\prime}}}&=\frac{A X_{\text{LO}}}{\sqrt{u\!+\!V_{\text{RIN}}}} \left(\sqrt{\eta_{\text{d}}} \hat{x}_{M}+\sqrt{1-\eta_{\text{d}}} \hat{x}_{{v_1}}\right)\\&+\frac{X_{\text{el}}+X_{\text{RIN}}}{\sqrt{u+V_{\text{RIN}}}}.
\end{split}
\end{equation}

For ease of calculation, we denote $u^{\prime}$ as $u^{\prime} = mu$, where $m = 1 + V_{\text{RIN}}/u$. Due to Gaussian variables $X_{\text{el}}$ and $X_{\text{RIN}}$ immune from Eve, they can be written as $\sqrt{V_{\text{el}}}\hat{x}_{{v_2}}$ and $\sqrt{V_{\text{RIN}}}\hat{x}_{{v_3}}$ respectively, with the variable $\hat{x}_{{v_2}}$ and $\hat{x}_{{v_3}}$ of variance 1 representing for vacuum noise. Then equation (\ref{appendixequ:2}) can be derived by
\begin{equation}
\begin{aligned}
x_{\text{ori}}^{\prime} &\!=\!\frac{A X_{\text{LO}}\!\left(\!\sqrt{\eta_{\text d}} \hat{x}_{M}\!+\!\sqrt{1\!-\!\eta_{\text d}} \hat{x}_{{v_1}}\!\right)\!+\!\sqrt{V_{\text{el}}}\hat{x}_{{v_2}}}{\sqrt{m u}}\!+\!\frac{\sqrt{V_{\text{RIN}}}}{\sqrt{m u}} \hat{x}_{{v_3}} \\
&\!=\!\sqrt{\frac{1}{m}}\!\left(\!\sqrt{\eta_{\text d}} \hat{x}_{M}\!+\!\sqrt{1\!-\!\eta_{\text d}} \hat{x}_{{v_1}}\!+\!\sqrt{v_{\text{el}}} \hat{x}_{{v_2}}\!\right)\!\!+\!\sqrt{1\!-\!\frac{1}{m}} \hat{x}_{{v_3}} \\
&\!=\!\sqrt{\frac{1}{m}} x_{\text{ori}}\!+\!\sqrt{1\!-\!\frac{1}{m}} \hat{x}_{{v_3}}.
\end{aligned}
\end{equation}

It is straightforward to see that the measurement result of monitoring deviates from the expected variable $x_{\text{ori}}$ to the realistic variable $x_{\text{ori}}^{\prime}$. At this time if Alice and Bob have no realization of this deviation, they will use $x_{\text{ori}}$ instead of $x_{\text{ori}}^{\prime}$ in parameter estimation, which may be a security issue.

\section{Equivalences between EB scheme and P\&M scheme of one-time calibration method}\label{AppendixB}

As mentioned above, neglecting RIN leaves a security loophole and our proposed scheme can make sense to solve this thanks to OTC method. Here the equivalence establishment of the EB and the P\&M scheme of the monitoring module is given. Without loss of generality, the derivation of the equivalence in Alice's monitoring module is given for an example. As shown in Fig.~\ref{PM&EBscheme}, the variable of the quantum signal used for monitoring is $\hat{x}_{M_1}$, hence the measurement outcome of monitoring without normalization is

\begin{equation}
X_{M_3}\!=\!A X_{\text{LO}}\!\left(\!\sqrt{\eta_{\text{d}}} \hat{x}_{ {M_1}}\!+\!\sqrt{1-\eta_{\text{d}}} \hat{x}_{v_1}\!\right)\!\!+\!X_{\text{el}}\!+\!X_{\text{RIN}}.
\end{equation}

After normalized by the new SNU $u^{\prime}=u+V_{\text{el}}+V_{\text{RIN}}$, the measurement result of monitoring becomes
\begin{equation}
\begin{aligned}
x_{M_3}\!=\!\frac{X_{M_3}}{\sqrt{u^{{\prime}}}}&\!=\!\frac{A X_{\text{LO}}}{\sqrt{u\!+\!V_{\text {el}}\!+\!V_{\text{RIN}}}} \!\times\!\left(\sqrt{\eta_{\text{d}}} \hat{x}_{M_1}\!+\!\sqrt{1-\eta_{\text{d}}} \hat{x}_{{v}_1}\right)\\&\!+\!\frac{X_{\text {el}}\!+\!X_{\text{RIN}}}{\sqrt{u\!+\!V_{\text{el}}\!+\!V_{\text{RIN}}}}.
\end{aligned}
\end{equation}

In the sense that $X_{\text{el}}$ and $X_{\text{RIN}}$ are independent variables, their addition $X_{\text{el}} + X_{\text{RIN}}$ in the last term can be substituted with $\sqrt{V_{\text{el}} + V_{\text{RIN}}}\hat{x}_{v_2}$. Then $x_{M_3}$ is expressed as

\begin{equation}
\begin{aligned}
x_{M_3}&\!=\!\frac{A X_{\text{LO}}}{\sqrt{A^{2} X_{\text{LO}}^{2}\!+\!V_{\text{el}}\!+\!V_{\text{RIN}}}} \!\times\!\left(\sqrt{\eta_{\text{d}}} \hat{x}_{M_1}\!+\!\sqrt{1\!-\!\eta_{\text{d}}} \hat{x}_{{v}_1}\right) \\
&\!+\!\frac{\sqrt{V_{\text{el}}\!+\!V_{\text{RIN}}}}{\sqrt{A^{2} X_{\text{LO}}^{2}\!+\!V_{\text{el}}\!+\!V_{\text{RIN}}}} \hat{x}_{{v}_2}.
\end{aligned}
\end{equation}

Here we denote $\eta_{\text e} = \frac{A^2X^2_{\text{LO}}} {A^2X^2_{\text{LO}}+V_{\text{el}}+V_{\text{RIN}}}$, then the equation above can be modified by

\begin{equation}
x_{M_3}=\sqrt{\eta_{\text e}} \times\left(\sqrt{\eta_{\text{d}}} \hat{x}_{M_1}+\sqrt{1-\eta_{\text{d}}} \hat{x}_{{v}_1}\right)+\sqrt{1-\eta_{\text{e}}} \hat{x}_{{v}_2}.
\end{equation}

According to this equation, the equivalent EB scheme is established, where the finite detection efficiency and the electronic noise of the homodyne detector are imitated by two BSs with transmittance of $\eta_{\text d}$ and $\eta_{\text e}$ respectively. In particular, the BSs take the order of the one with transmittance $\eta_{\text d}$ and the one with transmittance $\eta_{\text e}$. In this EB scheme, the passive effect of the electronic noise of the practical homodyne detector together with the RIN of the laser can be observed and are reflected by the extra loss, in accordance with the only one-time operation in the P\&M scheme.

\section{Realistic Gaussian attack in practical scenario}\label{AppendixC}

The covariance matrix of the reduced state $\rho_{E_1E_2}$ of Eve's ancillary modes $E_1$ and $E_2$ which have quantum correlations with each other is

\begin{equation}
\gamma_{E_{1} E_{2}}=\left[\begin{array}{cc}
\omega_{A} \mathbf{I} & \mathbf{G} \\
\mathbf{G} & \omega_{B} \mathbf{I}
\end{array}\right] \text{with}~\mathbf{G}=\left[\begin{array}{ll}
g & 0 \\
0 & g^{\prime}
\end{array}\right],
\end{equation}
where $\omega_A$, $\omega_B$ are variances of the thermal noise introduced into the links by Eve. $g$, $g^{\prime}$ are correlated parameters which should satisfy specific physical restrictions, and $\mathbf{I}=\operatorname{diag}(1,1)$ is the identity matrix of 2 orders. After the interactions, all of Eve's ancillary modes are stored in the quantum memory and are measured at the end of the protocol. When $g \neq 0, g^{\prime} \neq 0$, the attack is so-called the two-mode attack, which is the most general eavesdropping approach for CV-MDI QKD. When $g = g^{\prime} = 0$, there are no correlations between the two entangling cloners, which means the attack is just a one-mode collective attack.

Here we investigate the negative EPR attack, the proved optimal correlated attack for CV-MDI QKD,
where the correlation between the two ancillary modes of Eve will destroy the Bell detection, and there exists the following relations

\begin{equation}
g^{\prime}=-g=\phi,
\end{equation}
where
\begin{equation}
\phi=\min \left\{\sqrt{\left(\omega_{A}-1\right)\left(\omega_{B}+1\right)}, \sqrt{\left(\omega_{A}+1\right)\left(\omega_{B}-1\right)}\right\}.
\end{equation}

\section{Derivation of covariance matrices}\label{AppendixD}
To compute the covariance matrices conditioned on Charlie's measurement outcomes, we give the results of
the needed elements in Eqs.~(\ref{Case1}), (\ref{Case2}) and (\ref{Case3}).

In \textbf{Case 1}, Alice and Bob's reduced covariance matrix is $\gamma_{B_1A_1FM} = \gamma_{B_1} \oplus \gamma_{A_1FM}$, with $\gamma_{B_1} = V \mathbf{I}$ and $\gamma_{A_1FM} = \gamma^{(A)}$, where $\gamma^{(*)}$ for each $* = A, B$ is derived as

\begin{widetext}
\begin{equation}
\gamma^{(\star)}=\left(\begin{array}{ccccc}
V \mathbf{I} & -\tau \zeta_{1} \mathbf{Z} & 0 \mathbf{I} & -\sqrt{\eta_{\star} T_{\mathrm{S}}} \zeta_{1} \mathbf{Z} & \sqrt{\eta_{\star}^{\prime} T_{\mathrm S}} \zeta_{1} \mathbf{Z} \\
-\tau \zeta_{1} \mathbf{Z} & \varphi \mathbf{I} & \sqrt{T_{\mathrm{S}}} \zeta_{2} \mathbf{Z} & \sqrt{\eta_{\star}} \delta \mathbf{I} & -\sqrt{\eta_{\star}^{\prime}} \delta \mathbf{I} \\
0 \mathbf{I} & \sqrt{T_{\mathrm{S}}} \zeta_{2} \mathbf{Z} & V_{\mathrm{S}} \mathbf{I} & -\sqrt{\eta_{\star}} \tau \zeta_{2} \mathbf{Z} & \sqrt{\eta_{\star}^{\prime}} \tau \zeta_{2} \mathbf{Z} \\
-\sqrt{\eta_{\star} T_{\mathrm{S}}} \zeta_{1} \mathbf{Z} & \sqrt{\eta_{\star}} \delta \mathbf{I} & -\sqrt{\eta_{\star}} \tau \zeta_{2} \mathbf{Z} & \left(\eta_{\star} \sigma+1\right) \mathbf{I} & -\sqrt{k} \eta_{\star}^{\prime} \sigma \mathbf{I} \\
\sqrt{\eta_{\star}^{\prime} T_{\mathrm{S}}} \zeta_{1} \mathbf{Z} & -\sqrt{\eta_{\star}^{\prime} \delta \mathbf{I}} & \sqrt{\eta_{\star}^{\prime}} \tau \zeta_{2} \mathbf{Z} & -\sqrt{k} \eta_{\star}^{\prime} \sigma \mathbf{I} & \left(\eta_{\star}^{\prime} \sigma+1\right) \mathbf{I}
\end{array}\right),
\end{equation}
\end{widetext}
where
\begin{equation}
\begin{aligned}
\zeta_{1} &:=\sqrt{V^{2}-1}, \\
\zeta_{2} &:=\sqrt{V_{\mathrm{S}}^{2}-1}, \\
\eta_{A} &:=\eta_{\mathrm{d}} \eta_{\mathrm{e}}\left(1-T_{M}\right), \\
\eta_{A}^{\prime} &:=\left(1-\eta_{\mathrm{d}}\right) \eta_{\mathrm{e}}\left(1-T_{M}\right), \\
\eta_{B} &:=\eta_{\mathrm{d}} \eta_{\mathrm{e}}\left(1-T_{K}\right), \\
\eta_{B}^{\prime} &:=\left(1-\eta_{\mathrm{d}}\right) \eta_{\mathrm{e}}\left(1-T_{K}\right), \\
\tau &:=\sqrt{1-T_{\mathrm{S}}}, \\
k &:=\eta_{\mathrm{d}} /\left(1-\eta_{\mathrm{d}}\right), \\
\varphi &:=\left(1-T_{\mathrm{S}}\right) V+T_{\mathrm{S}} V_{\mathrm{S}}, \\
\delta &:=\sqrt{T_{\mathrm{S}}\left(1-T_{\mathrm{S}}\right)}\left(V-V_{\mathrm{S}}\right),\\
\sigma&:=T_{\mathrm{S}}V+\left(1-T_{\mathrm{S}}\right)V_{\mathrm{S}}-1,
\end{aligned}
\end{equation}
and $\mathbf{Z}=\operatorname{diag}(1,-1)$ is Pauli matrix.

Similarly in \textbf{Case 2}, Alice and Bob's reduced covariance matrix is $\gamma_{A_1B_1GK} = \gamma_{A_1} \oplus \gamma_{B_1GK}$, with $\gamma_{A_1} = V \mathbf{I}$ and $\gamma_{B_1GK} = \gamma^{(B)}$. In \textbf{Case 3}, Alice and Bob's reduced covariance matrix is $\gamma_{A_1FMB_1GK} = \gamma_{A_1FM} \oplus \gamma_{B_1GK} = \gamma^{(A)} \oplus \gamma^{(B)}$.

The covariance matrix $\mathbf{R}$ of the relay's outcomes has the form
\begin{equation}
\mathbf{R}=\left(\begin{array}{cc}
\frac{1}{2} \theta & 0 \\
0 & \frac{1}{2} \theta^{\prime}
\end{array}\right),
\end{equation}
where $\theta$, $\theta^{\prime}$ are
\begin{equation}
\begin{aligned}
\theta &=\left(\eta_{A} T_{M}+\eta_{B} T_{K}\right) \sigma+\left(\eta_{A}+\eta_{B}\right)+\lambda, \\
\theta^{\prime}&=\left(\eta_{A} T_{M}+\eta_{B} T_{K}\right) \sigma+\left(\eta_{A}+\eta_{B}\right)+\lambda^{\prime} .
\end{aligned}
\end{equation}
with
\begin{equation}
\begin{aligned}
\lambda=\left(1-\eta_{A}\right) \omega_{A}&+\left(1-\eta_{B}\right) \omega_{B}-2 g \sqrt{\left(1-\eta_{A}\right)\left(1-\eta_{B}\right)}, \\
\lambda^{\prime} =\left(1-\eta_{A}\right) \omega_{A}&+\left(1-\eta_{B}\right) \omega_{B}+2 g^{\prime} \sqrt{\left(1-\eta_{A}\right)\left(1-\eta_{B}\right)}.
\end{aligned}
\end{equation}

This covariance matrix only consists of the variances of $x_C$ and $p_D$, and is dependent on the total transmittance of each links. By substituting the conditions $T_K = 1$, $T_M = 1$ and $T_M = T_K \neq 1$ respectively, we can derive the covariance matrices $\mathbf{R}^{(1)}$, $\mathbf{R}^{(2)}$ and $\mathbf{R}^{(3)}$ in three cases. Finally, the covariance matrices representing for the correlations between the trusted modes and modes $C$ and $D$ in three cases are derived as
\begin{equation}
\begin{gathered}
\mathbf{C}^{(1)}=\left(\begin{array}{c}
-\sqrt{\frac{1}{2} T_{B} T_{\mathrm{S}}} \zeta_{1} \mathbf{I} \\
\sqrt{\frac{1}{2} T_{A} T_{M}} \mathbf{C}^{(A)}
\end{array}\right), \\
\mathbf{C}^{(2)}=\left(\begin{array}{c}
\sqrt{\frac{1}{2} T_{A} T_{\mathrm{S}}} \zeta_{1} \mathbf{Z} \\
-\sqrt{\frac{1}{2} T_{B} T_{K}} \mathbf{C}^{(B)}
\end{array}\right),
\end{gathered}
\end{equation}
and
\begin{equation}
\mathbf{C}^{(3)}=\left(\begin{array}{c}
\sqrt{\frac{1}{2} T_{A} T_{M}} \mathbf{C}^{(A)} \\
-\sqrt{\frac{1}{2} T_{B} T_{K}} \mathbf{C}^{(B)}
\end{array}\right),
\end{equation}
where $\mathbf{C}^{(A)}$ and $\mathbf{C}^{(B)}$ take the following form respectively
\begin{equation}
\mathbf{C}^{(A)}=\left(\begin{array}{c}
\sqrt{T_{\mathrm S}} \zeta_{1} \mathbf{Z} \\
-\delta \mathbf{I} \\
\tau \zeta_{2} \mathbf{Z} \\
-\sqrt{\eta_{\star}} \sigma \mathbf{I} \\
\sqrt{\eta_{\star}^{\prime}} \sigma \mathbf{I}
\end{array}\right),
\end{equation}
and
\begin{equation}
\mathbf{C}^{(B)}=\left(\begin{array}{c}
\sqrt{T_{\mathrm S}} \zeta_{1} \mathbf{I} \\
-\delta \mathbf{Z} \\
\tau \zeta_{2} \mathbf{I} \\
-\sqrt{\eta_{\star}} \sigma \mathbf{Z} \\
\sqrt{\eta_{\star}^{\prime}} \sigma \mathbf{Z}
\end{array}\right).
\end{equation}

\section{Parameter estimation of CV-MDI QKD with the proposed scheme}\label{AppendixE}

Here the detailed parameter estimation process in \textbf{Case 1} is given for an example, and the parameters used to calculate the secret key rate in \textbf{Case 2} and \textbf{Case 3} can be estimated in a similar way.

As shown in the protocol procedure, the excess noise will be evaluated with the public measurement results $\left \{x_C, p_D\right \}$ and the quadratures of the prepared states $\left\{x_{A_1}, p_{A_1}\right\}$ and $\left\{x_{B_1},p_{B_1}\right\}$. Firstly, Alice and Bob will modified the data as:
\begin{equation}
\begin{aligned}
x_1&=x_{A_1}-x_{B_1},\\
p_2&=p_{A_1}+p_{B_1}.
\end{aligned}
\end{equation}
Then the quadrature of the modified prepared states $x_1$, $p_2$ and the measurement results $x_C$, $p_D$ are linked through the following relation
\begin{equation}
\begin{aligned}
y_1=t_1x_1+z_1,\\
y_2=t_2p_2+z_2,
\end{aligned}
\end{equation}
which represent for that the channel is normal linear. $z_1$, $z_2$ follow centered normal distribution with unknown variance $\sigma_1^2 = 1 + t_1\varepsilon_1$, $\sigma_2^2 = 1 + t_2\varepsilon_2$, respectively. $\varepsilon_1$ and $\varepsilon_2$ are the channel excess noise for the two quantum channels linked to Charlie. Moreover $t_1=T_{\text S}T_M\eta_{A}$ is the total transmission efficiency of channels between Alice and Charlie, $t_2=T_{\text S}T_K\eta_B$ is the total transmission efficiency of channels between Bob and Charlie. The maximum-likelihood estimators for the normal linear model are known as
\begin{equation}
\begin{aligned}
\hat{t}_{1}&=\frac{\sum_{i=1}^{m} x_{1i} y_{1i}}{\sum_{i=1}^{m} x_{1i}^{2}},\\
\hat{t}_{2}&=\frac{\sum_{i=1}^{m} x_{2i} y_{2i}}{\sum_{i=1}^{m} x_{2i}^{2}},\\
\hat{\sigma}^{2}_{1}&=\frac{1}{m} \sum_{i=1}^{m}\left(y_{1i}-\hat{t}_1 x_{1i}\right)^{2},\\
\hat{\sigma}^{2}_{2}&=\frac{1}{m} \sum_{i=1}^{m}\left(y_{2i}-\hat{t}_2 x_{2i}\right)^{2}.
\end{aligned}
\end{equation}

Meanwhile, based on the prepared states and monitoring data, we have
\begin{equation}
\begin{aligned}
\left\langle x_{A_{1}}, x_{M(K)_3}\right\rangle&=-\sqrt{T_{\mathrm S} \eta_{\text e} \eta_{\text d}\left(1-T_{M(K)}\right)}\sqrt{V^2-1},\\
\left\langle p_{A_{1}}, p_{M(K)_3}\right\rangle&=\sqrt{T_{S} \eta_{e} \eta_{d}\left(1-T_{M(K)}\right)} \sqrt{V^2-1},\\
\left\langle x_{M(K)_3}^{2}\right\rangle&=\left\langle p_{M(K)_3}^{2}\right\rangle
\\&=T_{\mathrm S} \eta_{\text e} \eta_{\text d}\left(1-T_{M(K)}\right)\left(V-1+\varepsilon_{\text S}\right)+1.
\end{aligned}
\end{equation}
Now the source noise $\varepsilon_{\text S}$ can be estimated. However, other parameters are related to each other, so their estimates are unable to attained directly. The effective way to solve this is to scan the possible values of one of the parameters. At this time, the values of other parameters also change, and the values of the parameters that make the secret key rate reach lower bound are selected. Here the values of these parameters can also be further limited to meet the requirements of actual conditions.

\end{appendix}

~\\
~\\
~\\


\end{document}